\documentclass[a4paper,12pt]{article}
\pdfoutput=1
\usepackage{jheppubnew}
\usepackage{pgf,tikz,pgfplots}
\usetikzlibrary{cd}
\pgfplotsset{compat=1.15}
\usepackage{mathrsfs}
\usetikzlibrary{arrows}
\usepackage{dsfont}
\usepackage{amsfonts}
\usepackage{amsmath,amssymb}
\usepackage{physics}
\usepackage{amsthm}
\usepackage{setspace}
\usepackage{relsize}
\usepackage{todonotes}
\usepackage{xpatch}
\usepackage[none]{hyphenat}
\usetikzlibrary{matrix}

\usepackage{tikz-cd}

\xpatchcmd\section{\large}{\Large}{}{}
\xpatchcmd\subsection{\normalsize}{\large}{}{}
\xpatchcmd\subsubsection{\normalsize}{\normalsize}{}{}

\newcommand{\CH}{\mathcal{H}}
\newcommand{\CD}{\mathcal{D}}

\newcommand{\CB}{\mathcal{B}}
\newcommand{\CC}{\mathcal{C}}

\newcommand{\CT}{\mathcal{T}}

\newcommand{\CZ}{\mathcal{Z}}

\newcommand{\DZ}{\mathds{Z}}
\newcommand{\trl}{\mathbf{1}}
\newcommand{\rvline}{\hspace*{-\arraycolsep}\vline\hspace*{-\arraycolsep}}

\newcommand{\be}{\begin{equation}}
\newcommand{\ee}{\end{equation}}
\newcommand{\bea}{\begin{eqnarray}}
\newcommand{\eea}{\end{eqnarray}}

\newcommand{\FC}{\mathfrak{C}}
\newcommand{\FD}{\mathfrak{D}}

\def \Q{\mathbb{Q}}
\def \Z{\mathbb{Z}}

\def \bock{\mathcal{B}}

\def \ra{\mathrm{a}}
\def \rb{\mathrm{b}}
\def \rc{\mathrm{c}}
\def \rd{\mathrm{d}}

\newtheorem{theorem}{Theorem}[section]

\newtheorem{prop}{Proposition}[section]
\newtheorem{corollary}{Corollary}[section]

 \title{Braidings on topological operators, anomaly of higher-form symmetries and the SymTFT}
\author{Pavel Putrov}
\author{and Rajath Radhakrishnan}
\affiliation{International Centre for Theoretical Physics,\\ Strada Costiera 11, Trieste 34151, Italy}

\abstract{The anomaly of non-invertible higher-form symmetries is determined by the braiding of topological operators implementing them. In this paper, we study a method to classify braidings on topological line and surface operators by leveraging the fact that topological operators which admit a braiding are symmetries of their associated SymTFT. This perspective allows us to formulate an algorithm to explicitly compute all possible braidings on a given fusion category, bypassing the need to solve the hexagon equations. Additionally, using 3+1d SymTFTs, we determine braidings on various fusion 2-categories. We prove a necessary and sufficient condition for the fusion 2-categories $\Sigma \CC$, 2Vec$_G^{\pi}$ and Tambara-Yamagami (TY) 2-categories TY$(A,\pi)$ to admit a braiding.}

\begin{document}

\maketitle

%\flushbottom
\newpage

\section{Introduction}

One of the powerful handles that we have to constrain the RG flows of QFTs is the 't Hooft anomaly of symmetries. For invertible 0-form symmetries, the anomaly of these symmetries can be classified. For example, in 1+1d, the anomaly of $G$ 0-form symmetry, where $G$ is a finite group, acting on a bosonic QFT is classified by $H^2(G,U(1))$ \cite{Kapustin:2014zva}. The modern understanding of symmetries is in terms of topological operators \cite{Gaiotto:2014kfa}. These generalized symmetries are described by mathematical structures richer than a group. Unlike for groups, the classification of these generalized symmetries is inherently tied to the classification of their anomalies (see the reviews \cite{Schafer-Nameki:2023jdn,Brennan:2023mmt,Shao:2023gho}). The anomaly of a non-invertible symmetry is part of the data of a higher fusion category describing the topological operators implementing it \cite{Frohlich:2009gb,Bhardwaj:2017xup,Bhardwaj:2022yxj,Copetti:2023mcq}. For example, in 1+1d, a non-invertible symmetry $\CC$ implemented by topological line operators with fusion rules
\be
a \times b= \sum_{c\in \CC} N_{ab}^c~c~,
\ee
is described by a fusion category structure on this fusion ring.\footnote{We also need the fusion category to have a spherical structure. We will assume that the QFT is unitary in which case the fusion category is expected to be unitary. A unitary fusion category has a unique spherical structure \cite[Proposition 9.5.1]{etingof2016tensor}.}. The anomaly of this symmetry is an obstruction to gauging this symmetry and is determined by the associator or $F$ symbols of the fusion category \cite{Frohlich:2009gb,Bhardwaj:2017xup}. Therefore, the classification of anomaly of non-invertible 0-form symmetries requires a classification of $F$ symbols on a given fusion ring. While this also has a cohomological answer (see, for example, \cite[Section 7.22]{etingof2016tensor}), the explicit classification of $F$ symbols for fusion rings is hard for an arbitrary ring. In fact, we do not even have a general method to determine whether a given fusion ring admits a fusion category structure on it.\footnote{For unitary fusion categories there are strong necessary conditions for categorification which can be used to rule out candidate fusion rings \cite{Liu_2021,etingof2023necessaryconditionunitarycategorification}.}

In this paper, we will consider the classification of anomalies for non-invertible higher-form symmetries in 2+1d and 3+1d. In 2+1d, 1-form symmetry is implemented by topological line operators. The line operators and its anomaly is described by a braided fusion category $\CC$.\footnote{Since the line operators also have spin, the braided fusion category must have more structure making it a premodular category. However, a unitary braided fusion category is spherical, and is automatically premodular \cite[Proposition A.4]{henriques2015categorified} (see also \cite{222098}).} In particular, consider the $S$-matrix which captures the non-trivial linking between two line operators (see Fig. \ref{fig:S matrix}). If the $S$-matrix is non-trivial, then the lines operators cannot be simultaneously gauged \cite{Gaiotto:2014kfa,Hsin:2018vcg}. Therefore, for gauging $\CC$ we must have 
\be
\label{eq:1form anomaly vanishing condition}
S_{a,b}:= \frac{d_ad_b}{D} ~ \forall a,b \in \CC~. 
\ee
where $d_a$ is the quantum dimension and $D=\sqrt{\sum_{a\in \CC}d_a^2}$ is a normalization factor. 
\begin{figure}[h!]
    \centering

\tikzset{every picture/.style={line width=0.75pt}} %set default line width to 0.75pt        

\begin{tikzpicture}[x=0.75pt,y=0.75pt,yscale=-1,xscale=1]

%Curve Lines [id:da4710837499938114] 
\draw    (374.64,149.64) .. controls (356.86,136.28) and (360.5,105) .. (385.75,99.75) ;
%Curve Lines [id:da4305725330253384] 
\draw    (351.31,160.33) .. controls (403.5,158.13) and (399.24,114.04) .. (378.9,104.8) ;
%Curve Lines [id:da5533231041676764] 
\draw    (351.31,160.33) .. controls (295.5,157) and (310.5,83) .. (371.17,100.82) ;
%Curve Lines [id:da30675669770260106] 
\draw    (385.75,99.75) .. controls (456.5,83) and (454.64,172.81) .. (383.53,155.88) ;

% Text Node
\draw (308,126) node    {$a$};
% Text Node
\draw (450,127) node    {$b$};
% Text Node
\draw (235,118.4) node [anchor=north west][inner sep=0.75pt]    {$S_{ab} =$};

\end{tikzpicture}
    \caption{Double braiding between two line operators is captured by the $S$-matrix.}
    \label{fig:S matrix}
\end{figure}
If this condition is satisfied, then $\CC$ is a symmetric braided fusion category, which is known to be equivalent to Rep$(G)$ for some finite group $G$ \cite{deligne2002categories}.\footnote{The condition on $S_{a,b}$ implies that lines must be bosonic or fermionic. Throughout this paper, we will assume that the lines which braid trivially with all other lines are bosonic. If there are fermionic lines, then $G$ is a super group.}  In this case, the 1-form symmetry can indeed be gauged and the dual symmetry is the invertible 0-forms symmetry $G$. Therefore, the anomaly of the 1-form symmetry, which is an obstruction to gauging it is captured by the non-triviality of the S-matrix. The $S$-matrix is determined by the $R$ symbols and fusion rules of $\CC$. 
\be
S_{ab}=\frac{1}{D} \sum_c d_c \text{Tr}(R_{\bar b a}^cR_{a\bar b}^c)~.
\ee
Therefore, to classify the anomaly of $\CC$ we need to find all consistent choices of $R$ symbols on it. For a fusion category $\CC$, the allowed braidings on it can be classified by solving the hexagon equations. However, similar to the classification problem for $F$ symbols, this is a very complicated problem for general fusion rings. In fact, looking at the hexagon equations, it is not even obvious which fusion categories admit a braiding. Moreover, since the $R$ symbols depend on a choice of basis of the fusion spaces, the solutions to the hexagon equations must be classified up to basis transformations. 

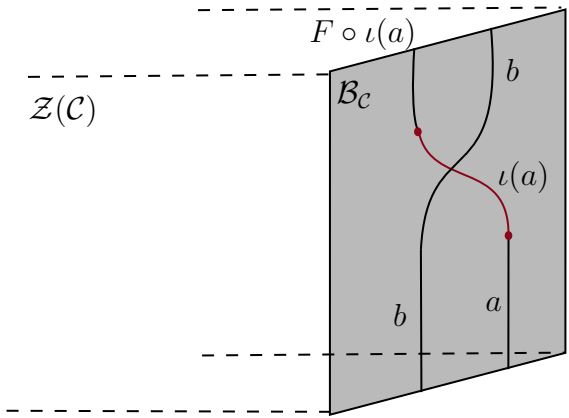
\begin{figure}[h!]
    \centering

\tikzset{every picture/.style={line width=0.75pt}} %set default line width to 0.75pt        

\begin{tikzpicture}[x=0.75pt,y=0.75pt,yscale=-1,xscale=1]

%Shape: Parallelogram [id:dp30241267080257395] 
\draw  [color={rgb, 255:red, 0; green, 0; blue, 0 }  ,draw opacity=1 ][fill={rgb, 255:red, 74; green, 74; blue, 74 }  ,fill opacity=0.38 ] (355.91,240) -- (355.94,67.29) -- (473.5,36.15) -- (473.47,208.87) -- cycle ;
%Curve Lines [id:da988440307380812] 
\draw [color={rgb, 255:red, 139; green, 6; blue, 24 }  ,draw opacity=1 ]   (400.23,97.65) .. controls (406.5,127) and (445.5,111) .. (445.07,149.88) ;
%Straight Lines [id:da018007275006587498] 
\draw    (401.59,228.64) -- (401.33,157.05) ;
%Straight Lines [id:da29508419290330234] 
\draw  [dash pattern={on 4.5pt off 4.5pt}]  (205.18,70.07) -- (360.54,68.76) ;
%Straight Lines [id:da7037040175103887] 
\draw  [dash pattern={on 4.5pt off 4.5pt}]  (290.07,36.5) -- (471.44,35.18) ;
%Straight Lines [id:da8454549995972991] 
\draw  [dash pattern={on 4.5pt off 4.5pt}]  (292.11,210.19) -- (473.47,208.76) ;
%Straight Lines [id:da25867664469226814] 
\draw  [dash pattern={on 4.5pt off 4.5pt}]  (194.5,238) -- (355.92,238.53) ;
%Straight Lines [id:da12223042371553872] 
\draw    (444.99,217.08) -- (445.07,149.88) ;
%Curve Lines [id:da8702526649725029] 
\draw    (401.33,157.05) .. controls (404.5,96) and (443.5,130) .. (436.32,45.42) ;
%Curve Lines [id:da2565306656413251] 
\draw    (398.05,55.66) .. controls (396.95,72.04) and (396.95,92.53) .. (400.23,97.65) ;
%Shape: Ellipse [id:dp7602602691923276] 
\draw  [color={rgb, 255:red, 139; green, 6; blue, 24 }  ,draw opacity=1 ][fill={rgb, 255:red, 139; green, 6; blue, 24 }  ,fill opacity=1 ] (443.66,149.88) .. controls (443.66,148.92) and (444.29,148.15) .. (445.07,148.15) .. controls (445.84,148.15) and (446.47,148.92) .. (446.47,149.88) .. controls (446.47,150.84) and (445.84,151.61) .. (445.07,151.61) .. controls (444.29,151.61) and (443.66,150.84) .. (443.66,149.88) -- cycle ;
%Shape: Ellipse [id:dp5976271768403947] 
\draw  [color={rgb, 255:red, 139; green, 6; blue, 24 }  ,draw opacity=1 ][fill={rgb, 255:red, 139; green, 6; blue, 24 }  ,fill opacity=1 ] (398.42,97.65) .. controls (398.42,96.69) and (399.05,95.92) .. (399.83,95.92) .. controls (400.6,95.92) and (401.23,96.69) .. (401.23,97.65) .. controls (401.23,98.6) and (400.6,99.38) .. (399.83,99.38) .. controls (399.05,99.38) and (398.42,98.6) .. (398.42,97.65) -- cycle ;

% Text Node
\draw (204.12,77.8) node [anchor=north west][inner sep=0.75pt]    {$\CZ( \CC)$};
% Text Node
\draw (385.96,182.94) node [anchor=north west][inner sep=0.75pt]    {$b$};
% Text Node
\draw (432.11,179.54) node [anchor=north west][inner sep=0.75pt]    {$a$};
% Text Node
\draw (344.22,39.14) node [anchor=north west][inner sep=0.75pt]    {$F\circ \iota ( a)$};
% Text Node
\draw (357.94,70.69) node [anchor=north west][inner sep=0.75pt]    {$\CB_{\CC}$};
% Text Node
\draw (438.04,111.06) node [anchor=north west][inner sep=0.75pt]    {$\iota ( a)$};
% Text Node
\draw (442.23,59.35) node [anchor=north west][inner sep=0.75pt]    {$b$};

\end{tikzpicture}
    \caption{Braiding the line $a$ with $b$ through the bulk using the maps $\iota$ and $F$.}
    \label{fig:braiding through the bulk}
\end{figure}

In this paper, we will outline an alternate approach to classifying braidings on a fusion category, circumventing the need to solve the hexagon equations. We will use the fact that if a fusion category of line operators admits a braiding, then it must be symmetries of its SymTFT \cite[Section 7]{MUGER2003159}.\footnote{The $d$+1 dimensional SymTFT has been crucial to compute anomalies and understand various aspects of generalized symmetries in $d$ dimensional QFTs \cite{Freed:2012bs,Gaiotto:2020iye,Kong:2020cie,Apruzzi:2021nmk,Freed:2022qnc,Chatterjee:2022kxb,Kaidi:2022cpf,Kaidi:2023maf,Zhang:2023wlu,Bhardwaj:2023fca,Bhardwaj:2023idu,Bhardwaj:2023bbf,Putrov:2024uor,lu2024exploring,Antinucci:2024ltv,copetti2024defect,bhardwaj2024boundary,etxebarria2024symtft}. Our work continues this direction to compute the possible anomalies of higher-form symmetries. A crucial difference to the analysis of 0-form symmetries using SymTFTs is that we study anomaly of a higher-form symmetry in $d$ dimensions by embedding it in the SymTFT of the same dimension.} Consider the SymTFT $\CZ(\CC)$ of a fusion category $\CC$ which has a canonical boundary condition $\CB_{\CC}$ on which the line operators from the category $\CC$. Two lines $a,b$ on $\CB_{\CC}$ cannot braid with each other on the boundary itself. However, we can define a braiding on them by bringing the line $a$ to the bulk and then taking it back onto the boundary on the other side of line $b$. Consider the bulk-to-boundary map (monoidal functor) $F:\CZ(\CC) \to \CC$, which is given by bringing the bulk lines onto the boundary.\footnote{In category theory, $F$ is known as the forgetful functor \cite[Section 7.13]{etingof2016tensor}.} Assume that there is a boundary-to-bulk map $\iota: \CC \to \CZ(\CC)$ which we use to ``push" the boundary line $a$ into the bulk for moving it around $b$. Taking the line $a$ around $b$ through the bulk using $\iota$ and $F$ defines a map (see Fig. \ref{fig:braiding through the bulk})
\be
a \times b \to b \times  F\circ \iota(a)~.
\ee
For this to be a braiding between $a$ and $b$, we need $F\circ \iota(a)=a$ for all $a\in \CC$. Note that the bulk-to-boundary map is not invertible. However, the above condition implies that $F$ must be the inverse of $\iota$ when acting on $\iota(\CC)$. Therefore,  for $\CC$ to admit a braiding, there should be an embedding $\iota$ of $\CC$ in $\CZ(\CC)$.\footnote{More precisely, $F$ is a monoidal functor from $\CZ(\CC)$ to $\CC$. $F$ is required to be invertible on $\iota(\CC)$, and $\iota$ is the inverse of $F$. This implies that $\iota$ is injective on objects and morphisms. Therefore, it is an embedding of $\CC$ in $\CZ(\CC)$.} This is a sufficient condition for $\CC$ to admit a braiding because $\iota(\CC)$ is a subcategory of line operators in the SymTFT in which all lines have a braiding. However, the embedding $\iota$ may not be unique, and in general different embeddings specify different braidings on $\CC$. Therefore, the anomaly of $\CC$ can be classified by embedding $\CC$ in its SymTFT. Combining this with the result that the braiding of line operators in a TQFT is completely fixed by its modular data \cite{ng2024recoveringrsymbolsmodulardata}, we will outline an algorithm to compute all braidings on a fusion category.

In 3+1d, the higher-form symmetries are implemented by topological line and surface operators. Since topological line operators braid trivially in 3+1d, the 2-form symmetry implemented by them is anomaly free. However, the 1-form symmetry implemented by surface operators can be anomalous. Moreover, there can be a mixed-anomaly between the 1-form and 2-form symmetry captured by the braiding of surface operators with line operators. The topological surface and line operators of QFTs form a braided fusion 2-category.\footnote{Braided monoidal 2-categories were introduced in \cite[Appendix B]{schommer2009classification}. For a modern introduction to these categories, see \cite[Section 12.1]{johnson20202}.} Classifying the possible anomalies of 1-form symmetry and its mixed anomaly with the 2-form symmetry requires a classification of braidings on a fusion 2-category. Similarly to the story in 2+1d, we will understand this problem using the 3+1d SymTFT. A fusion 2-category $\FC$ admits a braiding if and only if it can be embedded inside its SymTFT $\CZ(\FC)$. We will show that $\FC$ admits a braiding only if $\CZ(\FC)$ is a Dijkgraaf-Witten theory in 3+1d. We will use the surface and line operator content of this TQFT and their fusion rules to study embedding of $\FC$ in it for various important special cases of fusion 2-categories. For the fusion 2-category 2Vec$_G^{\pi}$, where $\pi \in Z^4(G,U(1))$  we show that (Theorem \ref{th:braidings on 2VecG} and Corollary \ref{cor:2VecG})

\vspace{0.2cm}
\textit{2Vec$_G^{\pi}$ admits a braiding if and only if $G$ is abelian and $[\pi]$ is trivial in $H^4(G,U(1))$.}
\vspace{0.2cm}

\noindent As a consequence of this result, we argue that any QFT with a $G$ 1-form symmetry necessarily contains (non-invertible) 0-form symmetries implemented by condensation defects of the 1-form symmetry. For the Tambara-Yamagami fusion 2-categories TY$(A,\pi)$ determined by a finite abelian group $A$ and 4-cocycle $\pi \in Z^4((A\times A) \rtimes \DZ_2,U(1))$, we show that (Theorem \ref{th: braiding on TY} and Corollary \ref{cor:2TY})

\vspace{0.2cm}
\textit{TY$(A,\pi)$ admits a braiding if and only if $A\cong \DZ_2^M$ for some integer $M$ and $[\pi]$ is trivial in $H^4((A\times A) \rtimes \DZ_2,U(1))$.}
\vspace{0.2cm}

The structure of this paper is as follows. In Section \ref{sec:braidings on fusion} we will describe the classification of braidings on line operators using the SymTFT. We will start with a brief review of SymTFTs. In Section \ref{sec: computing modular data}, we will explain how to compute the modular data of $\CZ(\CC)$ directly from the fusion category $\CC$. We will also explain how to identify the line operators in $\CZ(\CC)$ which can end on the canonical gapped boundary $\CB_{\CC}$ of $\CZ(\CC)$. We end the section with the steps involved in explicitly computing the braidings on a fusion category. In Section \ref{sec:1catexamples}, we give various examples to illustrate this method. In Section \ref{sec:braidings on 2cat}, we move on to studying braidings on topological surfaces and lines. We review the properties of the 3+1d SymTFT of a fusion 2-category. We also review various properties of the 3+1d Dijkgraaf-Witten theory in Section \ref{sec: SymTFT for G}. We end this section by describing how to determine the braidings on a fusion 2-category by embedding it in its SymTFT. In Section \ref{sec:examples 2cat}, we illustrate our method in various examples. Some of the technical calculations involved in these examples are moved to the appendices. We conclude with a discussion of possible future directions. 

\section{Braidings on topological lines operators from 2+1d SymTFT}

\label{sec:braidings on fusion}

\subsection{2+1d SymTFT}

Let $\CC$ be a braided fusion category. The SymTFT of $\CC$ is a 2+1d TQFT described by the Drinfeld centre $\CZ(\CC)$ \cite{Freed:2012bs,Gaiotto:2020iye,Kong:2020cie,Apruzzi:2021nmk,Freed:2022qnc,Chatterjee:2022kxb,Kaidi:2022cpf,Kaidi:2023maf}. Throughout this paper, we will denote the line operators in $\CC$ by $a,b,c$ and the line operators in $\CZ(\CC)$ as $x,y,z$. In fact, every line operator in $\CZ(\CC)$ is of the form
\be
(a,e_a)
\ee
where $e_a(b) : a\times b \to b\times a$ is the half-braiding.  Note that even if $\CC$ does not admit a braiding, the line operators in $\CZ(\CC)$ can braid with each other. However, if $\CC$ admits a braiding, then we can embed $\CC$ in $\CZ(\CC)$ as follows 
\bea 
\label{eq:embedding}
&&\iota: \CC \hookrightarrow~~\CZ(\CC)~, \\
&& \hspace{0.5cm} a \mapsto (a,R_{\_ ,a}^{-1})~. 
\eea
Physically, the embedding $\iota$ can be understood as follows. There exists a canonical Lagrangian algebra $L_{\CC}$ and a corresponding gapped boundary $\CB_{\CC}$ on which the line operators form the unitary fusion category $\CC$. If $\CC$ admits a braiding then $\iota$ allows us to bring a line operator from the boundary $\CB_{\CC}$ into the bulk.\footnote{When $\CC$ does not admit a braiding, then there is no such map which can be used to map a simple line operator on the boundary into a simple line operator in the bulk. However, there exists a boundary-to-bulk map called the Induction functor \cite[Section 9.2]{etingof2016tensor}. This map necessarily takes simple line operators on the boundary to \textit{non-simple} line operators in the bulk.} Similarly, we have the bulk-to-boundary map 
\bea
\label{eq:forgetful}
&& F: \CZ(\CC) \to \CC~, \\
&& \hspace{0.5cm} (a,e_a) \mapsto a ~.
\eea
which determines the fusion of a bulk line operator on the gapped boundary $\CB_{\CC}$. In particular, we have that
\be
\iota \circ F : \CC \to \CC
\ee
is the identity map on $\CC$. $\CZ(\CC)$ admits a canonical gapped boundary $\CB_{\CC}$ on which the line operators form the category $\CC$. The set of line operators which can end on this gapped boundary forms a Lagrangian algebra $L_{\CC}$. Using the bulk-to-boundary map $F$, the simple line operators $(a,e_a)$ in $L_{\CC}$ are precisely those such that $a$ contains the identity line $\trl$. We can also identify the other Lagrangian algebras in $\CZ(\CC)$ using the method described in \cite{Putrov:2024uor}. Let $\CB$ be a fusion category such that 
\be
\CZ(\CB)\cong \CZ(\CC)~.
\ee
We will assume that the algebra objects in $\CB$ are known. Using, \cite[Theorem 4.1]{Putrov:2024uor}, we know that for every Lagrangian algebra $L$ in $\CZ(\CB)$, $F(L)$ is of the form
\be
\label{eq:F(L)}
F(L)=\sum_{A_m} A_m~,
\ee
where the R.H.S is a sum of a full set of Morita equivalence class of algebras in $\CB$. Therefore, using the knowledge of the R.H.S, in many cases, we can completely fix the Lagrangian algebra objects in $\CZ(\CC)\cong \CZ(\CB)$. See \cite[Section 6.3]{Putrov:2024uor} for explicit examples.

\subsection{Computing the modular data of $\CZ(\CC)$}

\label{sec: computing modular data}

\begin{figure}[h!]
    \centering

\tikzset{every picture/.style={line width=0.75pt}} %set default line width to 0.75pt        

\begin{tikzpicture}[x=0.75pt,y=0.75pt,yscale=-1,xscale=1]

%Straight Lines [id:da32400869264876087] 
\draw    (322,49) -- (322,186) ;
%Curve Lines [id:da23873857251959685] 
\draw    (317,228) .. controls (270,228) and (273,117) .. (323,105) ;
%Curve Lines [id:da5825970196548762] 
\draw    (322,146) .. controls (374,175) and (339,233) .. (317,228) ;
%Shape: Ellipse [id:dp05504443620363275] 
\draw  [color={rgb, 255:red, 0; green, 0; blue, 0 }  ,draw opacity=1 ][fill={rgb, 255:red, 0; green, 0; blue, 0 }  ,fill opacity=1 ] (319.83,106) .. controls (319.83,104.88) and (320.54,103.97) .. (321.42,103.97) .. controls (322.29,103.97) and (323,104.88) .. (323,106) .. controls (323,107.12) and (322.29,108.03) .. (321.42,108.03) .. controls (320.54,108.03) and (319.83,107.12) .. (319.83,106) -- cycle ;
%Shape: Ellipse [id:dp4774411238452486] 
\draw  [color={rgb, 255:red, 0; green, 0; blue, 0 }  ,draw opacity=1 ][fill={rgb, 255:red, 0; green, 0; blue, 0 }  ,fill opacity=1 ] (320,146) .. controls (320,144.88) and (320.71,143.97) .. (321.58,143.97) .. controls (322.46,143.97) and (323.17,144.88) .. (323.17,146) .. controls (323.17,147.12) and (322.46,148.03) .. (321.58,148.03) .. controls (320.71,148.03) and (320,147.12) .. (320,146) -- cycle ;
%Shape: Ellipse [id:dp5441840997644659] 
\draw  [color={rgb, 255:red, 0; green, 0; blue, 0 }  ,draw opacity=1 ][fill={rgb, 255:red, 0; green, 0; blue, 0 }  ,fill opacity=1 ] (320.42,188.03) .. controls (320.42,186.91) and (321.13,186) .. (322,186) .. controls (322.87,186) and (323.58,186.91) .. (323.58,188.03) .. controls (323.58,189.15) and (322.87,190.06) .. (322,190.06) .. controls (321.13,190.06) and (320.42,189.15) .. (320.42,188.03) -- cycle ;

% Text Node
\draw (324,52.4) node [anchor=north west][inner sep=0.75pt]    {$a$};
% Text Node
\draw (324,119.9) node [anchor=north west][inner sep=0.75pt]    {$b$};
% Text Node
\draw (324,156.4) node [anchor=north west][inner sep=0.75pt]    {$c$};
% Text Node
\draw (354,188.4) node [anchor=north west][inner sep=0.75pt]    {$d$};

\end{tikzpicture}
    \caption{Lasso action corresponding to the tube algebra basis element $\CT_{abc}^d$.}
    \label{fig:tube algebra action}
\end{figure}
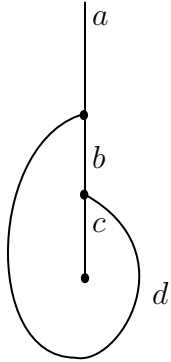

The modular $S$ and $T$ matrices of the SymTFT $\CZ(\CC)$ will play a crucial role in our analysis. While computing all the data of the MTC $\CZ(\CC)$ is complicated, its $S$ and $T$ matrices can be computed using the data of the fusion category $\CC$ through a straightforward algorithm. Following  \cite{Williamson:2017uzx}, we outline the procedure to compute it from the tube algebra of $\CC$. To ease the notation, we will assume that the fusion spaces of $\CC$ are at most $1$-dimensional. The tube algebra Tube$(\CC)$ has a basis of elements of the form
\be
\CT_{abc}^d~.
\ee
 It denotes the lasso action of the line operator $d$ on a twisted sector operator of the line $c$ transforming it into a twisted sector operator of line $a$ (see Fig. \ref{fig:tube algebra action}).
The multiplication of the tube algebra is the composition of lasso actions. This is completely determined by the $F$ symbols of the category $\CC$.  The tube algebra multiplication is given by
\be
\CT_{abc}^d \CT_{a'b'c'}^{d'}= v_d \delta_{c,a'} \sum_{e,f} \sqrt{\frac{d_d d_{d'}}{d_{f}}} B_{b'\bar d}^{e} (B_{bd}^{c})^{*}[F^{b}_{d'b'\bar d}]_{ce} ([F_{ed'd}^{c'}]_{fb'})^{*} ([F_{dd'e}^{a}]_{bf})^{*} \CT_{c'ea}^{f}~.
\ee
 where $[F^{b'}_{dbd'}]_{ce}$ are the components of the $F$ symbols of $\CC$ in a basis, $v_a:=\frac{[F_{a\bar a a}^a]_{\trl \trl}}{|[F_{a\bar a a}^a]_{\trl \trl}|}$ is the Frobenius-Schur indicator of $a$, $d_a:= \frac{1}{[F_{a\bar a a}^{a}]_{\trl \trl}}$ is the quantum dimension of $a$, $B_{ab}^c:=\frac{d_ad_b}{d_c} [F_{ab\bar b}^{a}]_{c\trl}$. It is known that the simple line operators of the SymTFT $\CZ(\CC)$ are in one-to-one correspondence with the irreducible representations of this algebra \cite{evans1995ocneanu,izumi2000structure,MUGER2003159}. Computing these irreducible representations explicitly is equivalent to computing the half-braidings required to define line operators of the Drinfeld centre.\footnote{For example, see \cite[Section 5.1.1]{Choi:2024tri}.} These irreducible representations are in one-to-one correspondence with indecomposable central idempotents of Tube$(\CC)$. Let $P_x$ be such an idempotent. Note that we use the same notation for the subscript of the idempotents as well as the simple objects of $\CZ(\CC)$ as they are in one-to-one correspondence. We have a basis transformation
\be
P_x = \sum_{a,b,c,d\in \CC} M_{x,abcd}~ \CT_{abc}^d~.
\ee
By definition, they satisfy
\be
P_x P_y = \delta_{x,y} P_x~,~ P_x \CT_{abc}^d= \CT_{abc}^d P_x~,~ \text{ and } P_x\neq P_y + P_z~ .
\ee
These conditions can be turned into a set of constraints on the basis transformation matrices $M_{x,abcd}$ which can, in turn, be turned into a constructive polynomial time algorithm to determine them as described in \cite[Appendix C]{Bultinck:2015bot}\cite{friedl1985polynomial}. While the explicit description of the irreducible representations of Tube$(\CC)$ in terms of matrices requires a choice of basis, the $M_{x,abcd}$ contain basis-independent information about these representations. Indeed, the $M_{x,abcd}$ are the characters of the irreducible representations of the tube algebra.\footnote{This is true for all semi-simple algebras. See, for example, \cite[Theorem 3.8]{ram1991representation}.}\footnote{The characters of tube algebra and their orthogonality relations were recently studied in \cite{Choi:2024tri}.}  Note that as in the representation theory of finite groups, it is easier to compute $M_{x,abcd}$ than the explicit irreducible representations of the tube algebra. 
The modular data of $\CZ(\CC)$ is fully determined by $M_{x,abcd}$. The action of $S$ and $T$ on the tube algebra basis is given by 
\be
S(\CT_{abc}^d)=\delta_{ac} \sum_{e} v_{d} B_{e\bar d}^{a} B_{da}^{e} (B_{bd}^a)^{*} (A_{ad}^{e})^{*} ([F_{da\bar d}^{a}]_{eb})^{*} \CT_{ded}^{\bar a} ~~\text{ and } ~~ T(\CT_{abc}^d)=\CT_{a\trl a}^p \CT_{abc}^d~.
\ee
where $A_{ab}^{c}:=\sqrt{\frac{d_ad_b}{d_c}} ([F_{\bar a a b}^{b}]_{\trl c})^{*}$. These arise from the $S$ and $T$ transformation on a 2-torus supporting the tube algebra basis element $\CT_{abc}^d$~. The modular $S$ and $T$ matrix of the SymTFT is obtained from computing this action on the indecomposable central idempotents. The components of the modular data matrices of the SymTFT are given by \cite{Williamson:2017uzx} 
\be
S_{xy}= \frac{D_y \text{Tr}(P_y^{\dagger}S(P_x))}{D_{x}\text{Tr}(P_y^{\dagger}P_y)}~,~ T_{xy}=\delta_{xy} \frac{D_y \text{Tr}(P_y^{\dagger}T(P_x))}{D_x\text{Tr}(P_y^{\dagger}P_y)}~.
\ee
Here, $P_x$ must be thought of as a column vector in the basis $T_{abc}^d$ with components $M_{x,abcd}$. $S(P_x)$ is also a column vector and $P_y^{\dagger}S(P_x)$ is a matrix. $D_x$ is the dimension of the subalgebra of Tube$(\CC)$ determined by the projector $P_x$. 

Along with the modular date of $\CZ(\CC)$, the discussion in the following section will crucially depend on the simple line operators which form the canonical Lagrangian algebra $L_{\CC}$ of $\CZ(\CC)$. These are the line operators which can end on the canonical gapped boundary of $\CZ(\CC)$ hosting the line operators $\CC$ on it. Recall that a line operator in $\CZ(\CC)$ if of the form $(a,e_a)$. The bulk-to-boundary map is
\be
F((a,e_a))=a~.
\ee
The line operators which can end on the canonical gapped boundary are those such that $F((a,e_a))$ contains the identity line. Therefore, for this to be true the generically non-simple line $a$ must contain $\trl$. In this section, we also learned that the simple line operators in the SymTFT are in one-to-one correspondence with the irreducible representations of the tube algebra. Under this correspondence, a representation corresponding to $(a,e_a)$ acts on the Hilbert space $\CH_a$. $\CH_a$ is the direct sum of twisted Hilbert spaces for the simple line operators contained in the decomposition of the generically non-simple line $a$. Therefore, the line operators $(a,e_a)$ in $L_{\CC}$ correspond to representations of the tube algebra which act non-trivially on the untwisted Hilbert space along with potentially other twisted Hilbert spaces. Such representations can be readily read-off from the expression of the idempotents $P_x$ introduced above. 

As an example, consider the Fibonacci fusion category with line operators $\trl, \tau$. The tube algebra has the basis
\be
\CT_{\trl\trl\trl}^\trl~,~ \CT_{\tau \tau \tau}^\trl~,~ \CT_{\trl \tau \trl}^\tau~,~ \CT_{\trl \tau \tau}^\tau~,~ \CT_{\tau \trl \tau}^\tau~,~ \CT_{\tau \tau \trl}^\tau~,~ \CT_{\tau \tau \tau}^\tau~.
\ee
The idempotents are given by \cite[Appendix D.1]{Bultinck:2015bot}
\be
\begin{split}
P_1&= \frac{1}{\sqrt{5}} \bigg(\frac{1}{\phi} \CT_{\trl\trl\trl}^\trl + \sqrt{\phi} \CT_{\trl\tau\trl}^{\tau} \bigg)~,\\
P_2&= \frac{1}{\sqrt{5}} \bigg(\frac{1}{\phi} \CT_{\tau\tau\tau}^\trl + \frac{1}{\sqrt{\phi}} e^{-\frac{4 \pi i}{5}}\CT_{\tau\trl\tau}^{\tau} + e^{\frac{3 \pi i}{5}} \CT_{\tau\tau\tau}^\tau \bigg)~,\\
P_3&= \frac{1}{\sqrt{5}} \bigg(\frac{1}{\phi} \CT_{\tau\tau\tau}^\trl + \frac{1}{\sqrt{\phi}} e^{\frac{4 \pi i}{5}}\CT_{\tau\trl\tau}^{\tau} + e^{-\frac{3 \pi i}{5}} \CT_{\tau\tau\tau}^\tau \bigg)~,\\
P_4&= \frac{1}{\sqrt{5}} \bigg(\phi \CT_{\trl\trl\trl}^\trl + \CT_{\tau\tau\tau}^\trl -\sqrt{\phi} \CT_{\trl\tau\trl}^\tau + \sqrt{\phi} \CT_{\tau \trl \tau}^\tau + \frac{1}{\phi} A_{\tau\tau\tau}^\tau \bigg)~.
\end{split}
\ee
From this, we can determine that the idempotents $P_1$ and $P_4$ correspond to simple line operators in the canonical Lagrangian algebra. This is because they contain tube algebra basis elements $\CT_{abc}^d$ with $c=\trl$.

\subsection{Classifying braidings on topological line operators}

Let $\CC$ be a fusion category. Let us assume that $\CC$ admits a braiding. Recall that we have an embedding of $\CC$ in $\CC(\CZ)$ given by the map $\iota$ in \eqref{eq:embedding}. Consider the subcategory $\iota(\CC)$ in $\CZ(\CC)$. It is clear that $\CC$ has the following properties
\be
\text{dim}(\iota(\CC))= \text{dim}(\CC), ~ \text{ and } F(\iota(\CC))=\CC~.
\ee
where dim$(\CC):=\sum_a d_a^2$. This is true for any choice of braiding on $\CC$. Conversely, let $\CZ(\CC)$ be the SymTFT of a fusion category $\CC$. Let $\CD$ be a fusion subcategory of $\CZ(\CC)$ with the following properties
\be
\label{eq:Dconditions}
\text{dim}(\CD)=\text{dim}(\CC)~, \text{ and } O(\CD) \cap O(L_{\CC}) =\{\trl\}~.
\ee
 where $O(\CD)$ is the set of simple line operators of $\CD$ and $O(L_{\CC})$ is the set of simple line operators in the decomposition of $L_{\CC}$ into simple line operators. The second condition above denotes that there are no non-trivial line operators common to both $\CD$ and the canonical Lagrangian algebra $L_{\CC}$. In \cite{nikshych2018classifying}, it was shown that these conditions imply
\be
F(\CD)\cong \CC
\ee
as fusion categories. Indeed, $\text{dim}(\CD)=\text{dim}(\CC)$ implies that $F:\CC \to \CD$ is surjective and $O(\CD) \cap O(L_{\CC}) =\{\trl\}$ implies that it is injective. This shows that any $\CD$ which satisfies \eqref{eq:Dconditions} is the image of an embedding $\iota$ of $\CC$ in $\CZ(\CC)$. Moreover, $F\circ \iota$ is the identity map of $\CC$. 

Since $\CD$ is a fusion subcategory of $\CZ(\CC)$, it is braided. Therefore, $F$ equips $\CC$ with a braiding given by the braiding of line operators in $\CD$. If there is no $\CD$ which satisfies the conditions in \eqref{eq:Dconditions}, then $\CC$ does not admit a braiding.  Note that classifying such fusion subcategories $\CD$ only requires the knowledge of the fusion ring of $\CZ(\CC)$ and the Lagrangian algebra object $L_{\CC}$. Crucially, we do not need to know the $F$ and $R$ matrices of $\CZ(\CC)$ or the explicit multiplication of the algebra $L_{\CC}$. This provides a simple method to count the number of distinct braidings on a fusion category. Indeed, \cite{nikshych2018classifying} contains explicit expressions parametrizing the embedding of $\CC$ in $\CZ(\CC)$ for various fusion categories.

If $\CC$ admits some braiding, to determine it explicitly, we need to compute the $R$ matrices of $\CC$ in some basis. This is also the $R$ matrices for line operators in $\CD$. The latter can be comptuted using the remarkable result in \cite{ng2024recoveringrsymbolsmodulardata}, that the $R$ matrices of a modular category is completely determined by its modular data! Let $x,y,z,\dots$ be the simple line operators in $\CZ(\CC)$. Consider an arbitrary complete ordering on the simple objects. There exists a basis in which the $R$ matrices of $\CZ(\CC)$ are given by 
\be
\label{eq:Rmatrix}
R_{xy}^z = \begin{cases} I_{N_{xy}^{z}} & \mbox{if }  x>y~, \vspace{0.2cm} \\  \vspace{0.2cm}
			\frac{\theta_z}{\theta_x\theta_y} I_{N_{xy}^z} & \mbox{if }  x<y~,\\
			\frac{\sqrt{\theta_z}}{\theta_x} \Delta_{x}^z & \mbox{if }  x=y~.
			\end{cases} 
\ee
where $T_{xy}=\delta_{x,y}\theta_x$, $\Delta_x^z$ is a $N_{xx}^z \times N_{xx}^z$ matrix of the form
\be
\Delta_x^z=\begin{pmatrix}
  I_{d_{+}} & \rvline & 0 \\
\hline
  0 & \rvline & -I_{d_{-}}
\end{pmatrix}~,
\ee
and $I_{n}$ is the identity matrix of integer $n$. The integers $d_{\pm}$ are given by 
\be
\label{eq:d}
d_{\pm} = \frac{1}{2} \bigg (N_{xx}^z \pm \frac{1}{\sqrt{\theta_z} \text{dim}(\CZ(\CC))} \sum_{v,w \in \CZ(\CC)} d_{v} N_{vw}^{x} \bar S_{z,w} \frac{\theta_v^2}{\theta_w^2} \bigg )~,
\ee

In summary, the classification of braidings (if any) on $\CC$ proceeds as follows:
\begin{enumerate}
	\item Find the modular data of $\CZ(\CC)$ and the canonical Lagrangian algebra $L_{\CC}$ of $\CZ(\CC)$ corresponding to $\CC$ following the discussion in section \ref{sec: computing modular data}.
	\item Let $\Sigma$ be the set of fusion subcategories $\CD$ of $\CZ(\CC)$ satisfying
	\be
\text{dim}(\CD)=\text{dim}(\CC)~, \text{ and } 	\CD \cap L_{\CC} = \{\trl\}~, 
	\ee 
	If $\Sigma$ is empty, then $\CC$ does not admit any braiding. Each elemnt of $\Sigma$ corresponds to a braiding on $\CC$ and all braidings on $\CC$ are captured in this way. 
	\item For every $\CD \in \Sigma$, we get a braiding on $\CC$ with the explicit $R$ matrices given by 
	\be
\label{eq:R matrix for embedding}
	R|_{\CD}
	\ee
 	where $R$ is the braiding matrix of $\CZ(\CC)$ which can be computed using the modular data of $\CZ(\CC)$ and equations \eqref{eq:Rmatrix},\eqref{eq:d}.
\end{enumerate}
Note that the procedure above only requires the modular data of the SymTFT $\CZ(\CC)$, which can in turn be determined from the $F$ symbols of the fusion category. Determining the modular data of $\CZ(\CC)$ is particularly simple if there exists a Morita equivalent fusion category $\CB$ for which the modular data of $\CZ(\CB)$ is known. In this case, we have an equivalence of categories
\be
\phi: \CZ(\CB) \to \CZ(\CC)~,
\ee
using which we can determine the braidings on $\CC$. Moreover, if the algebra objects in $\CB$ are known, in many cases, we can use \eqref{eq:F(L)} to determine the braidings on all fusion categories Morita equivalent to $\CB$. 

Once we choose a braiding on $\CC$, it specifies the $S$ matrix which gives the anomaly of the 1-form symmetry $\CC$. More generally, the anomaly of a set of line operators which are not closed under fusion is determined by a lack of commutative algebra structure on them. We briefly outline how braidings on $\CC$ affect the set of commutative algebras in $\CC$ in Appendix \ref{ap: anomaly of lines not closed under fusion}.

\section{Examples}

\label{sec:1catexamples}

\subsection{Braidings on $\text{Vec}_G^{\omega}$}

The fusion category Vec$_{G}^{\omega}$ for a finite group $G$ and a  3-cocycle $\omega \in Z^3(G,U(1))$ captures the symmetry $G$ and anomaly $\omega$ of a 0-form symmetry in a 1+1d QFT. If Vec$_{G}^{\omega}$ admits a braiding, then it can also act as the symmetry of a 2+1d QFT. As we will demonstrate shortly, one can easily argue that the group $G$ must be abelian. However, this is not a sufficient condition of the existence of a braiding. The braidings on Vec$_G^{\omega}$ are well-studied \cite{joyal1993braided,drinfeld2010braidedfusioncategoriesi}. In this section, we will use the 2+1d SymTFT to arrive at the same result. The analysis in this section will be a precursor to analysing braidings on 2Vec$_G^{\pi}$.

The SymTFT for Vec$_G^{\omega}$ is the 2+1d twisted Dijkgraaf-Witten theory whose simple line operators can be labelled as
\be
([g],\pi_g) 
\ee
 where $[g]$ is a conjugacy class of $G$ and $\pi_g$ is an irreducible projective representation of the centralizer $C_g$ of $g \in G$ satisfying
\be
\label{eq:projective representations}
\pi_g(h) \pi_g(k)= \tau_g(\omega)(h,k) \pi_g(hk) ~ \forall h,k\in C_g~,   
\ee
where $\tau_g(\omega)(h,k)$ is the transgression of the 3-cocycle $\omega$ with respect to $g$ given by 
\be
\label{eq:transgression of 3-cocycle}
\tau_g(\omega)(h,k)= \frac{\omega(h,k,g)\omega(hkg(hk)^{-1},h,k)}{\omega(h,kgk^{-1},k)} ~.
\ee
The quantum dimension of these line operators is given by 
 \be
d_{([g],\pi_g)}= |[g]| \text{dim}(\pi_g)~.
\ee 
According to the general approach outlined in Section \ref{sec:braidings on fusion}, Vec$_{G}^{\omega}$  admits a braiding if and only if it can be embedded inside its SymTFT. Moreover the embedding $\iota:\mathrm{Vec}_G^\omega\hookrightarrow \mathcal{Z}(\mathrm{Vec}_G^\omega)$, must have the image $\CD:=\iota(\mathrm{Vec}_G^\omega)$ such that 
\be
O(\CD) \cap O(L_{\CC})= \{([e],\mathds{1})\}~,
\ee
where $e$ is the identity of $G$ and $\mathds{1}$ is its trivial irreducible representation. We have
\be
L_{\CC}= \sum_{\pi_g \in \text{Rep(G)}} ([e],\pi_g~)~.
\ee
Since all line operators in Vec$_{G}^{\omega}$ are invertible, we need some subset of invertible line operators in $\CZ(\text{Vec}_G^\omega)$ to form the group $G$. A simple line operator $([g],\pi_g)$ is invertible iff $d_{([g],\pi_g)}=1$. Therefore, we need $|[g]|=1$ and $\text{dim}(\pi_g)=1$. The former implies that $g$ must be in the center of $G$, while the latter requires that $\pi_g$ is equivalent to a linear 1-dimensional representation of $C_g$ which is the full group $G$ when $g$ is in the center. Clearly, the set of invertible line operators are closed under fusion and their fusion rule is given by 
\be
([g],\pi_g) \times ([h],\pi_h) = ([gh], \pi_g \otimes \pi_h)~.
\ee
For the embedding to respect the group multiplication of $G$, it must be given by some map of the form
\be
\iota: g \mapsto ([g],\sigma_g)~.
\ee
for some irreducible 1-dimensional representation $\sigma_g$ of $G$. Therefore, $|[g]|=1$ for all $g\in G$ implying that the group $G$ must be abelian. Moreover, using \eqref{eq:projective representations} and \eqref{eq:transgression of 3-cocycle} we have 
\be
\tau_g(\omega)(h,k)= \frac{\omega(h,k,g)\omega(g,h,k)}{\omega(h,g,k)} = d\psi_g(h,k) ~ \forall ~g\in G
\ee
for some 2-cochain $C^2(G,U(1))$ . 

\begin{theorem} 
\label{th:braidings on VecG}
Vec$_G^{\omega}$ admits a braiding if and only if $G$ is abelian and $\tau_g(\omega)=d\psi_g$ for some 2-cochain $\psi_g\in C^2(G,U(1))$ for all $g\in G$. 
\end{theorem}

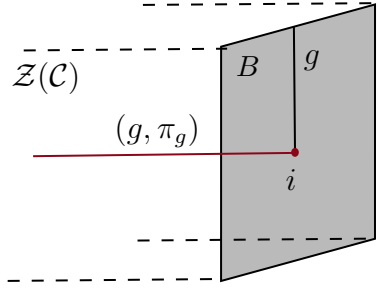
\begin{figure}[h!]
    \centering

\tikzset{every picture/.style={line width=0.75pt}} %set default line width to 0.75pt        

\begin{tikzpicture}[x=0.75pt,y=0.75pt,yscale=-1,xscale=1]

%Shape: Parallelogram [id:dp30241267080257395] 
\draw  [color={rgb, 255:red, 0; green, 0; blue, 0 }  ,draw opacity=1 ][fill={rgb, 255:red, 74; green, 74; blue, 74 }  ,fill opacity=0.38 ] (354.15,201.1) -- (354.17,83.24) -- (431,62) -- (430.98,179.86) -- cycle ;
%Straight Lines [id:da39765233905630826] 
\draw [color={rgb, 255:red, 139; green, 6; blue, 24 }  ,draw opacity=1 ]   (260.32,138.33) -- (391.06,136.67) ;
%Straight Lines [id:da018007275006587498] 
\draw    (391.06,136.67) -- (390.5,73) ;
%Straight Lines [id:da29508419290330234] 
\draw  [dash pattern={on 4.5pt off 4.5pt}]  (255.64,85.14) -- (357.17,84.25) ;
%Straight Lines [id:da7037040175103887] 
\draw  [dash pattern={on 4.5pt off 4.5pt}]  (315.14,62.24) -- (433.66,61.33) ;
%Straight Lines [id:da8454549995972991] 
\draw  [dash pattern={on 4.5pt off 4.5pt}]  (312.45,180.76) -- (430.98,179.78) ;
%Straight Lines [id:da25867664469226814] 
\draw  [dash pattern={on 4.5pt off 4.5pt}]  (247.62,201.07) -- (354.15,200.1) ;
%Shape: Ellipse [id:dp8653005467824513] 
\draw  [color={rgb, 255:red, 139; green, 6; blue, 24 }  ,draw opacity=1 ][fill={rgb, 255:red, 139; green, 6; blue, 24 }  ,fill opacity=1 ] (389.65,136.4) .. controls (389.65,135.44) and (390.28,134.67) .. (391.06,134.67) .. controls (391.83,134.67) and (392.46,135.44) .. (392.46,136.4) .. controls (392.46,137.36) and (391.83,138.13) .. (391.06,138.13) .. controls (390.28,138.13) and (389.65,137.36) .. (389.65,136.4) -- cycle ;

% Text Node
\draw (360.17,86.65) node [anchor=north west][inner sep=0.75pt]  [font=\small]  {$B$};
% Text Node
\draw (299.5,115.55) node [anchor=north west][inner sep=0.75pt]    {$( g,\pi _{g})$};
% Text Node
\draw (248,89.4) node [anchor=north west][inner sep=0.75pt]    {$\CZ( \CC)$};
% Text Node
\draw (394,83.4) node [anchor=north west][inner sep=0.75pt]    {$g$};
% Text Node
\draw (385,143.4) node [anchor=north west][inner sep=0.75pt]    {$i$};

\end{tikzpicture}
    \caption{The fusion of $([g],\pi) \in \CD$ on the gapped boundary $\CB_{\text{2Vec}_G^{\omega}}$ must produce the line operator $g\in \text{Vec}_G^{\omega}$. The junction vector space $V_{([g],\pi_g),g}$ is 1-dimensional with a basis vector $i$.}
    \label{fig:1cat fusion on boundary}
\end{figure}

The condition $\tau_{g}(\omega)=d\psi_g$ can be understood as the vanishing of an anomaly as follows. Consider the bulk-to-boundary map $F:\CZ(\CC) \to \CC$ which in this case is given by
\be
F(([g],\pi_g))= |[g]| \sum_{a \in [g]} a~.
\ee 
If $\iota$ is an embedding of Vec$_{G}^{\omega}$, we require $F|_{\CD} \circ \iota : \text{Vec}_{G}^{\omega} \to \text{Vec}_{G}^{\omega}$ to be the identity map. This is to ensure that $([g],\pi_g)\times ([h],\pi_h) \to ([h],\pi_h)\times F\circ \iota(([g],\pi_g))$ defines a braiding on $\text{Vec}_{G}^{\omega}$. Therefore, if $([g],\pi_g) \in \CD$, we must have (see Fig. \ref{fig:1cat fusion on boundary})
\be
F(([g],\pi_g)) = g~.
\ee
This makes it clear that $g$ must be in the center of $G$ and that the vector space of operators $V_{([g],\pi_g),g}$  at the junction of $([g],\pi_g)$ with the boundary must be 1-dimensional. $V_{([g],\pi_g),g}$ is a representation of the tube algebra of Vec$_G^{\omega}$ which has the multiplication \cite{Roche:1990hs,Bullivant:2019fmk,Bartsch:2023wvv} 
\be 
\CT_{g(hg)g}^h \CT_{g(kg)g}^k = ~ \tau_g(\omega)(h,k) ~\CT_{g(hkg)g}^{hk}~.
\ee
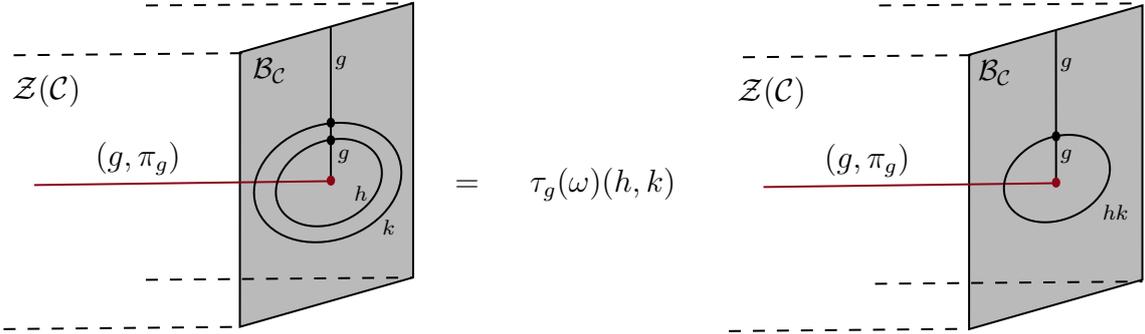
\begin{figure}[h!]
    \centering

\tikzset{every picture/.style={line width=0.75pt}} %set default line width to 0.75pt        

\begin{tikzpicture}[x=0.75pt,y=0.75pt,yscale=-1,xscale=1]
%uncomment if require: \path (0,285); %set diagram left start at 0, and has height of 285

%Shape: Parallelogram [id:dp30241267080257395] 
\draw  [color={rgb, 255:red, 0; green, 0; blue, 0 }  ,draw opacity=1 ][fill={rgb, 255:red, 74; green, 74; blue, 74 }  ,fill opacity=0.38 ] (164.14,220.83) -- (164.17,82.81) -- (250.66,57.95) -- (250.64,195.98) -- cycle ;
%Straight Lines [id:da39765233905630826] 
\draw [color={rgb, 255:red, 139; green, 6; blue, 24 }  ,draw opacity=1 ]   (61.5,149.5) -- (208.69,147.55) ;
%Straight Lines [id:da018007275006587498] 
\draw    (209.57,147.24) -- (209.5,70) ;
%Straight Lines [id:da29508419290330234] 
\draw  [dash pattern={on 4.5pt off 4.5pt}]  (51.23,84.21) -- (165.55,83.17) ;
%Straight Lines [id:da7037040175103887] 
\draw  [dash pattern={on 4.5pt off 4.5pt}]  (117.21,59.18) -- (250.66,58.12) ;
%Straight Lines [id:da8454549995972991] 
\draw  [dash pattern={on 4.5pt off 4.5pt}]  (117.19,197.18) -- (250.64,196.04) ;
%Straight Lines [id:da25867664469226814] 
\draw  [dash pattern={on 4.5pt off 4.5pt}]  (46.2,221.97) -- (166.14,220.83) ;
%Shape: Ellipse [id:dp8653005467824513] 
\draw  [color={rgb, 255:red, 139; green, 6; blue, 24 }  ,draw opacity=1 ][fill={rgb, 255:red, 139; green, 6; blue, 24 }  ,fill opacity=1 ] (207.98,147.24) .. controls (207.98,146.12) and (208.69,145.21) .. (209.57,145.21) .. controls (210.44,145.21) and (211.15,146.12) .. (211.15,147.24) .. controls (211.15,148.36) and (210.44,149.27) .. (209.57,149.27) .. controls (208.69,149.27) and (207.98,148.36) .. (207.98,147.24) -- cycle ;
%Shape: Ellipse [id:dp1530642965965604] 
\draw  [color={rgb, 255:red, 0; green, 0; blue, 0 }  ,draw opacity=1 ][fill={rgb, 255:red, 0; green, 0; blue, 0 }  ,fill opacity=1 ] (207.98,126.96) .. controls (207.98,125.84) and (208.69,124.93) .. (209.57,124.93) .. controls (210.44,124.93) and (211.15,125.84) .. (211.15,126.96) .. controls (211.15,128.08) and (210.44,128.99) .. (209.57,128.99) .. controls (208.69,128.99) and (207.98,128.08) .. (207.98,126.96) -- cycle ;
%Shape: Ellipse [id:dp8697669058588184] 
\draw  [color={rgb, 255:red, 0; green, 0; blue, 0 }  ,draw opacity=1 ][fill={rgb, 255:red, 0; green, 0; blue, 0 }  ,fill opacity=1 ] (207.95,118.15) .. controls (207.95,117.03) and (208.66,116.12) .. (209.53,116.12) .. controls (210.41,116.12) and (211.12,117.03) .. (211.12,118.15) .. controls (211.12,119.27) and (210.41,120.17) .. (209.53,120.17) .. controls (208.66,120.17) and (207.95,119.27) .. (207.95,118.15) -- cycle ;
%Shape: Parallelogram [id:dp6004568513928931] 
\draw  [color={rgb, 255:red, 0; green, 0; blue, 0 }  ,draw opacity=1 ][fill={rgb, 255:red, 74; green, 74; blue, 74 }  ,fill opacity=0.38 ] (528.14,222) -- (528.17,83.97) -- (614.66,59.12) -- (614.64,197.14) -- cycle ;
%Straight Lines [id:da4356442729548675] 
\draw [color={rgb, 255:red, 139; green, 6; blue, 24 }  ,draw opacity=1 ]   (425.5,150.5) -- (572.69,148.55) ;
%Straight Lines [id:da9411129328557802] 
\draw    (571.57,148.24) -- (571.5,72) ;
%Straight Lines [id:da21613190267715143] 
\draw  [dash pattern={on 4.5pt off 4.5pt}]  (415.23,85.21) -- (529.55,84.17) ;
%Straight Lines [id:da7492666028005923] 
\draw  [dash pattern={on 4.5pt off 4.5pt}]  (483.21,59.18) -- (614.66,58.12) ;
%Straight Lines [id:da9815317864689819] 
\draw  [dash pattern={on 4.5pt off 4.5pt}]  (481.19,199.18) -- (614.64,198.04) ;
%Straight Lines [id:da6996780597541282] 
\draw  [dash pattern={on 4.5pt off 4.5pt}]  (408.2,222.97) -- (528.14,221.83) ;
%Shape: Ellipse [id:dp17596105089294578] 
\draw  [color={rgb, 255:red, 139; green, 6; blue, 24 }  ,draw opacity=1 ][fill={rgb, 255:red, 139; green, 6; blue, 24 }  ,fill opacity=1 ] (569.98,148.24) .. controls (569.98,147.12) and (570.69,146.21) .. (571.57,146.21) .. controls (572.44,146.21) and (573.15,147.12) .. (573.15,148.24) .. controls (573.15,149.36) and (572.44,150.27) .. (571.57,150.27) .. controls (570.69,150.27) and (569.98,149.36) .. (569.98,148.24) -- cycle ;
%Shape: Ellipse [id:dp576520433648254] 
\draw  [color={rgb, 255:red, 0; green, 0; blue, 0 }  ,draw opacity=1 ][fill={rgb, 255:red, 0; green, 0; blue, 0 }  ,fill opacity=1 ] (569.98,124.96) .. controls (569.98,123.84) and (570.69,122.93) .. (571.57,122.93) .. controls (572.44,122.93) and (573.15,123.84) .. (573.15,124.96) .. controls (573.15,126.08) and (572.44,126.99) .. (571.57,126.99) .. controls (570.69,126.99) and (569.98,126.08) .. (569.98,124.96) -- cycle ;
%Shape: Ellipse [id:dp009083267406822415] 
\draw   (183.48,173.65) .. controls (168.34,164.79) and (167.03,146.24) .. (180.56,132.2) .. controls (194.1,118.17) and (217.35,113.97) .. (232.49,122.83) .. controls (247.63,131.68) and (248.94,150.24) .. (235.4,164.27) .. controls (221.87,178.31) and (198.62,182.5) .. (183.48,173.65) -- cycle ;
%Shape: Ellipse [id:dp903068459977244] 
\draw   (190.33,167.18) .. controls (179.75,160.99) and (179.33,147.5) .. (189.41,137.06) .. controls (199.48,126.61) and (216.22,123.16) .. (226.81,129.35) .. controls (237.39,135.54) and (237.8,149.03) .. (227.73,159.47) .. controls (217.66,169.92) and (200.91,173.37) .. (190.33,167.18) -- cycle ;
%Shape: Ellipse [id:dp09577703932299919] 
\draw   (553.75,165.15) .. controls (543.16,158.96) and (542.75,145.48) .. (552.82,135.03) .. controls (562.9,124.59) and (579.64,121.14) .. (590.22,127.32) .. controls (600.8,133.51) and (601.22,147) .. (591.15,157.44) .. controls (581.07,167.89) and (564.33,171.34) .. (553.75,165.15) -- cycle ;

% Text Node
\draw (169.17,84.37) node [anchor=north west][inner sep=0.75pt]  [font=\small]  {$\CB_{\CC}$};
% Text Node
\draw (90.5,126.29) node [anchor=north west][inner sep=0.75pt]    {$( g,\pi _{g})$};
% Text Node
\draw (48.89,92.5) node [anchor=north west][inner sep=0.75pt]    {$\CZ(\CC)$};
% Text Node
\draw (210.32,82.93) node [anchor=north west][inner sep=0.75pt]  [font=\scriptsize]  {$g$};
% Text Node
\draw (211.57,130.36) node [anchor=north west][inner sep=0.75pt]  [font=\scriptsize]  {$g$};
% Text Node
\draw (219.82,148.41) node [anchor=north west][inner sep=0.75pt]  [font=\scriptsize]  {$h$};
% Text Node
\draw (233.82,165.41) node [anchor=north west][inner sep=0.75pt]  [font=\scriptsize]  {$k$};
% Text Node
\draw (531.17,85.37) node [anchor=north west][inner sep=0.75pt]  [font=\small]  {$\CB_{\CC}$};
% Text Node
\draw (454.5,127.29) node [anchor=north west][inner sep=0.75pt]    {$( g,\pi _{g})$};
% Text Node
\draw (410.89,93.5) node [anchor=north west][inner sep=0.75pt]    {$\CZ(\CC)$};
% Text Node
\draw (572.32,83.93) node [anchor=north west][inner sep=0.75pt]  [font=\scriptsize]  {$g$};
% Text Node
\draw (572.32,129.93) node [anchor=north west][inner sep=0.75pt]  [font=\scriptsize]  {$g$};
% Text Node
\draw (593.32,157.93) node [anchor=north west][inner sep=0.75pt]  [font=\scriptsize]  {$hk$};
% Text Node
\draw (270,139.4) node [anchor=north west][inner sep=0.75pt]    {$=\ \ \ \ \tau _{g}( \omega )( h,k)$};

\end{tikzpicture}
    \caption{Composition of tube algebra action on the boundary line $g$.}
    \label{fig:lasso action on g}
\end{figure}

\noindent For a given $g$, this tube algebra is the twisted group algebra of $G$ with the 2-cocycle $\tau_g(\omega)(h,k)$. Therefore, $V_{([g],\pi_g),g}$ is a quantum mechanical system with global symmetry $G$ and anomaly $\tau_g(\omega)(h,k)$. Since all non-trivial projective representations of a group are higher-dimensional, $V_{([g],\pi_g),g}$ is 1-dimensional iff $\tau_g(\omega)(h,k)$ is trivial in cohomology for all $g \in G$.

Suppose Vec$_G^{\omega}$ satisfies Theorem \ref{th:braidings on VecG}. To classify braidings on it, we have to find all embeddings of Vec$_G^{\omega}$ in $\CZ(\text{Vec}_G^{\omega})$. All fusion subcategories of $\CZ(\text{Vec}_G^{\omega})$ were classified in \cite[Theorem 5.11]{naidu2009fusionsubcategoriesrepresentationcategories}. A fusion subcategory $\CC(K,H,B)$ of $\CZ(\text{Vec}_G^{\omega})$ is determined by two subgroups $K,H$ of $G$ and an $\omega$-bicharacter $B:K\times H\to U(1)$ which is a function $B:K\times H\to U(1)$ satisfying the constraints
\be
B(g,hk)=\tau_{g}(\omega)(h,k) B(g,h)B(g,k)~,~B(gh,k)= \tau_{k}(\omega)(g,h)B(g,k)B(h,k) ~.
\ee
Also, the dimension of this subcategory is dim$(\CC(K,H,B))=\frac{|K||G|}{|H|}$. A simple object in this subcategory is of the form 
\be
([k],\pi)~,
\ee
where $k\in K$ and $\pi$ is an irreducible projective representation of $G$ such that $\pi(h)=B(k,h)$ for all $h\in H$. We want $\CC(K,H,B)$ to be an embedding of Vec$_G^{\omega}$ in $\CZ(\text{Vec}_G^{\omega})$ if and only if it satisfies the conditions \eqref{eq:Dconditions}. Note that if $H$ is a proper normal subgroup of $G$, then there exists a non-trivial irreducible representation $\pi$ of $G$ such that Ker$(\pi)=H$. In this case, $([e],\pi)$ is a simple line operator in $\CC(K,H,B)$. Therefore, to satisfy 
\be
O(\CC(K,H,B))\cap O(L_{\CC})=([e],\mathds{1})~,
\ee
we must have $H=G$. Then, for
\be
\text{dim}(\CC(K,G,B))= |K|=|G|~,
\ee
we also must choose $K=G$. Therefore, we find that all embeddings of Vec$_G^{\omega}$ in $\CZ(\text{Vec}_G^{\omega})$ is given by $\CC(G,G,B)$ for some $\omega$-bicharacter $B$. The braiding of invertible line operators is completely determined by their topological spins \cite{drinfeld2010braidedfusioncategoriesi}. The topological spin of line operators in $\CC(G,G,B)$ is given by 
\be
\theta_{(k,\pi)}:=\pi(k)=B(k,k)~,
\ee
which determines a braiding on Vec$_G^{\omega}$. All braidings on Vec$_G^{\omega}$ arise in this way. 

\subsection{Braidings on Rep$(G)$}

Consider the fusion category Rep$(G)$. The SymTFT in this case is the $G$ Dijkgraaf Witten theory. The line operators of the SymTFT can be written as
\be
([g],\pi_g)~,
\ee
where $[g]$ is a conjugacy class of $G$ with representative $g$ and $\pi_g$ is an irreducible representation of the centralizer $C_g$ of $g$. Without loss of generality, we can choose the Lagrangian algebra corresponding to $\CC=\text{Rep}(G)$ to be
\be
L_{\CC}=\sum_{[g]\in C(G)} ([g],\mathds{1}_g) ~,
\ee
where $C(G)$ is the set of conjugacy classes of $G$ and $\mathds{1}_g$ is the trivial representation of $C_g$. Clearly, there is at least one embedding of Rep$(G)$ in the SymTFT with the image given by 
\be
\CD:= \sum_{\pi \in \text{Rep}(G)} ([e],\pi)~.
\ee
This implies the well known fact that the fusion category Rep$(G)$ admits a symmetric braiding for any $G$. Other embeddings of Rep$(G)$ can be classified in terms of two normal subgroups of $G$ and a non-degenerate $G$-invariant bilinear form \cite[Example 5.5]{nikshych2018classifying}. Since the modular data of Dijkgraaf-Witten theory is known \cite{Roche:1990hs,Coste:2000tq} we can use \eqref{eq:Rmatrix} and \eqref{eq:R matrix for embedding} to explicitly compute all braidings corresponding to these embeddings.  We now give some specific examples to illustrate this.

\subsubsection{$G=\DZ_2$}

Let us classify braidings on the fusion category Rep$(\DZ_2)$. We know that Rep$(\DZ_2)\cong$ Vec$_{\DZ_2}$, and the braidings on the latter are of course well known. Nevertheless, it will be a useful exercise to understand the steps outlined in the previous section. We know that the SymTFT in this case is the $\DZ_2$ Dijkgraaf-Witten theory with line operators given by 
\be
([e],\mathds{1}), ([e],\pi), ([g],\mathds{1}), ([g],\pi)~.
\ee
where $g$ is the generator of $\DZ_2$, $g^2=e$, and $\mathds{1},\pi$ are the two irreducible representations of $\DZ_2$. The canonical Lagrangian algebra for $\CC=\text{Rep}(\DZ_2)$ is 
\be
L_{\CC}= ([e],\mathds{1}) + ([g],\mathds{1})~.
\ee
Therefore, the set $\Sigma$ contains the fusion subcategories
\be
\CD_1=\{([e],\mathds{1}), ([e],\pi)\}~,~ \CD_2=\{([e],\mathds{1}),([g],\pi)\}~.
\ee
Clearly, both $\CD_1$ and $\CD_2$ are isomorphic to Rep$(\DZ_2)$ as fusion categories. Since $([e],\pi)$ is a bosonic line operator and $([g],\pi)$ is fermionic, $\CD_1$ corresponds to the trivial braiding on $\CC$ while $\CD_2$ corresponds to the non-trivial braiding.

\subsubsection{$G=\DZ_2 \times \DZ_2$}

Consider the Klein four-group 
\be
\DZ_2 \times \DZ_2=\langle g_1,g_2|g_1^2=g_2^2=(g_1g_2)^2=e\rangle~.
\ee
The SymTFT for Rep$(\DZ_2 \times \DZ_2)$ is the $\DZ_2 \times \DZ_2$ Dijkgraaf-Witten theory with line operators
\be
([g],\pi)
\ee
where $g \in \DZ_2 \times \DZ_2$ and $\pi \in \text{Rep}(\DZ_2 \times \DZ_2)=\{\mathds{1},\omega_1,\omega_2,\omega_1\omega_2\}$ are irreducible representations of $\DZ_2 \times \DZ_2$. A Lagrangian algebra corresponding to a gapped boundary with lines on it forming Rep$(\DZ_2 \times \DZ_2)$ is 
\be
L_{\CC}= ([e],\mathds{1})+([g_1],\mathds{1})+([g_2],\mathds{1})+([g_1g_2],\mathds{1})~.
\ee
Rep$_{\DZ_2 \times \DZ_2}$ can be embedded in the SymTFT orthogonal to the $L_{\CC}$ in many ways. The embedding is determined by assigning to $\omega_1$ and $\omega_2$ any two distinct non-trivial line operators in the SymTFT not contained in $L_{\CC}$. However, not all of these embeddings give us distinct braidings. For example, the embeddings
\be
\begin{split}
& \CD_1:=(e,\mathds{1})+([e],\omega_1)+([e],\omega_2)+([e],\omega_1\omega_2)~,\\
& \CD_2:=(e,\mathds{1})+([g_2],\omega_1)+([g_1],\omega_2)+([g_1g_2],\omega_1\omega_2)~,\\
\end{split}
\ee
both give the symmetric braiding on Rep$_{\DZ_2 \times \DZ_2}$. This is because there is a 0-form symmetry $\rho$ of the SymTFT such that $\rho(\CD_1)=\CD_2$.

\subsubsection{$G=S_3$}

Consider the group 
\be
S_3 = \{r,s | r^3 =s^2=e; srs=r^{-1}\}~.
\ee
The SymTFT of Rep$(S_3)$ is the $S_3$ Dijkgraaf-Witten theory. The line operators of the SymTFT along with their quantum dimensions and topological spins are given by the following table.
\begin{center}
\begin{tabular}{ |c|c|c|c|c|c|c|c|c|c|} 
\hline
 & $([e],\mathds{1})$ &$([e],\pi_1)$& $([e],\pi_2)$, & $([r], \mathds{1}_r)$& $([r],\omega)$  & $([r], \omega^2)$ &$([s],\mathds{1}_s)$ & $([s],\gamma)$ \\ \hline
$d_x$ & 1 & 1 & 2 & 2 & 2 & 2&3  &3\\ \hline
$\theta_{x}$ & 1 & 1 & 1 & 1 & exp$(\frac{2\pi i}{3})$ & exp$(\frac{-2\pi i}{3})$ & 1& -1\\ \hline
\end{tabular}
\end{center}
The modular S-matrix and fusion rules are given, for example, in \cite{Beigi:2010htr}. We have to choose a Lagrangian algebra such that the lines on the boundary form $\CC=\text{Rep}(S_3)$. We can choose
\be
L_{\CC}= ([e],\mathds{1}) + ([r],\mathds{1}_r) + ([s],\mathds{1}_s)~.
\ee
In this case, the set $\Sigma$ contains the following subcategories
\be
\CD_1= \{([e],\mathds{1}),([e],\pi_1),([e],\pi_2)\}~,~\CD_2= \{([e],\mathds{1}),([e],\pi_1),([r],\omega)\}~, \nonumber\ee
\be
\CD_3= \{([e],\mathds{1}),([e],\pi_1),([r],\omega^2)\}~.
\ee
$\CD_1$ corresponds to the trivial braiding on Rep$(S_3)$ while $\CD_2$ and $\CD_3$ corresponds to non-trivial braidings on Rep$(S_3)$. These latter two cases can all be understood as follows. Consider the TQFT $SU(3)_1$ Chern-Simons theory and its Galois conjugate. Both of these TQFTs have a charge conjugation symmetry. Gauging this symmetry without considering the twisted sector line operators results in line operators forming the fusion category Rep$(S_3)$ with non-trivial braidings.\footnote{Physically, such gaugings should be understood as coupling the TQFT to QFT such that the twisted-sector line operators become non-topological. Then, the twisted sectors do not contribute additional genuine topological lines on gauging the symmetry of the combined coupled system \cite{Balasubramanian:2024nei}.}

\subsection{Braiding on fusion categories admitting a modular braiding}

Suppose the fusion category $\CC$ admits a modular braiding with modular data given by $S$ and $T$. In this case, the SymTFT is 
\be
\CZ(\CC) \cong \CC \times \bar \CC
\ee
where $\bar \CC$ is the category with the opposite braiding. The canonical Lagrangian algebra corresponding to $\CC$ is
\be
L_{\CC}= \sum_{a\in \CC} ~ (a,\bar a)~.
\ee
To enumerate the possible distinct braidings on $\CC$, we have to classify all fusion subcategories of $\CZ(\CC)$ satisfying \eqref{eq:Dconditions} to find the set $\Sigma$. In fact, this has been done in \cite[Theorem 4.1]{nikshych2018classifying}. Clearly, if the braiding on $\CC$ is known, then the braiding on $\CZ(\CC)$ is easily obtained. Therefore, the braiding matrices $R|_{\CD}$ for any $\CD\in \Sigma$ are also easily obtained. Also, from the results in \cite{ng2024recoveringrsymbolsmodulardata}, we know that the braiding on $\CC$ can be determined in terms of its modular data. Therefore, we get the following result:

\begin{prop}
Let $\CC$ be a fusion category which admits a modular braiding with modular data $S,T$. Then all braidings on $\CC$ can be explicitly determined using $S,T$ alone.
\end{prop}

\subsection{Braidings on Tambara-Yamagami fusion category}

A Tambara-Yamagami fusion category TY$(A,\chi,\epsilon)$ is defined by a finite abelian group $A$, a non-degenerate symmetric bilinear form $\chi:A \times A\to U(1)$ and a choice of sign $\epsilon:=\pm 1\frac{1}{\sqrt{|A|}}$ \cite{tambara1998tensor}. The category TY$(A,\chi,\epsilon)$ describes invertible line operators $a \in A$ whose fusion is given by the group multiplication of $A$, and a non-invertible line operator $m$ whose fusion is given by  
\be
m \times m= \sum_{a\in A} a~.
\ee
The braidings on TY$(A,\chi,\epsilon)$ have been explicitly calculated in \cite{siehler2000braidedneargroupcategories}. In that paper, the author computes the braiding by explicitly solving the hexagon equations. In particular, they show that the a braiding exists if and only if $A\cong \DZ_2^M$ for some positive integer $M$. Following the examples in the previous sections, an alternate approach to arriving at the same result is by studying the embedding of TY$(A,\chi,\epsilon)$ in its SymTFT. This can be done using the data of the SymTFT $\CZ(\text{TY}(A,\chi,\epsilon))$ described in \cite{gelaki2009centersgradedfusioncategories} (see also \cite{Zhang:2023wlu}). In particular, the condition $A \cong \DZ_2^M$ can be obtained by studying the bulk-to-boundary map. 

\begin{figure}[h!]
    \centering

\tikzset{every picture/.style={line width=0.75pt}} %set default line width to 0.75pt        

\begin{tikzpicture}[x=0.75pt,y=0.75pt,yscale=-1,xscale=1]

%Shape: Parallelogram [id:dp2838200514736674] 
\draw  [color={rgb, 255:red, 0; green, 0; blue, 0 }  ,draw opacity=1 ][fill={rgb, 255:red, 74; green, 74; blue, 74 }  ,fill opacity=0.38 ] (342.15,214.1) -- (342.17,96.24) -- (419,75) -- (418.98,192.86) -- cycle ;
%Straight Lines [id:da4007412248772414] 
\draw [color={rgb, 255:red, 139; green, 6; blue, 24 }  ,draw opacity=1 ]   (256.32,151.33) -- (387.06,149.67) ;
%Straight Lines [id:da7722580120157283] 
\draw    (387.06,149.67) -- (386.5,83) ;
%Straight Lines [id:da45173727840687183] 
\draw  [dash pattern={on 4.5pt off 4.5pt}]  (243.64,98.14) -- (345.17,97.25) ;
%Straight Lines [id:da8969032791334047] 
\draw  [dash pattern={on 4.5pt off 4.5pt}]  (303.14,75.24) -- (421.66,74.33) ;
%Straight Lines [id:da01617691787406761] 
\draw  [dash pattern={on 4.5pt off 4.5pt}]  (300.45,193.76) -- (418.98,192.78) ;
%Straight Lines [id:da2070192645205432] 
\draw  [dash pattern={on 4.5pt off 4.5pt}]  (235.62,214.07) -- (342.15,213.1) ;
%Shape: Ellipse [id:dp21727107381925603] 
\draw  [color={rgb, 255:red, 139; green, 6; blue, 24 }  ,draw opacity=1 ][fill={rgb, 255:red, 139; green, 6; blue, 24 }  ,fill opacity=1 ] (385.65,149.4) .. controls (385.65,148.44) and (386.28,147.67) .. (387.06,147.67) .. controls (387.83,147.67) and (388.46,148.44) .. (388.46,149.4) .. controls (388.46,150.36) and (387.83,151.13) .. (387.06,151.13) .. controls (386.28,151.13) and (385.65,150.36) .. (385.65,149.4) -- cycle ;
%Straight Lines [id:da21623098652843797] 
\draw [color={rgb, 255:red, 139; green, 6; blue, 24 }  ,draw opacity=1 ]   (257,170) -- (406,169) ;
%Shape: Ellipse [id:dp8996263007726338] 
\draw  [color={rgb, 255:red, 139; green, 6; blue, 24 }  ,draw opacity=1 ][fill={rgb, 255:red, 139; green, 6; blue, 24 }  ,fill opacity=1 ] (405.65,168.4) .. controls (405.65,167.44) and (406.28,166.67) .. (407.06,166.67) .. controls (407.83,166.67) and (408.46,167.44) .. (408.46,168.4) .. controls (408.46,169.36) and (407.83,170.13) .. (407.06,170.13) .. controls (406.28,170.13) and (405.65,169.36) .. (405.65,168.4) -- cycle ;
%Straight Lines [id:da4967662812212267] 
\draw    (407.06,167.67) -- (407,77) ;
\draw (295.5,138.55) node [anchor=north west][inner sep=0.75pt]    {$x$};
% Text Node
\draw (200,102.4) node [anchor=north west][inner sep=0.75pt]    {$\CZ(\text{TY}(A,\chi,\epsilon))$};
% Text Node
\draw (369,93.4) node [anchor=north west][inner sep=0.75pt]    {$m$};
% Text Node
\draw (391,93.4) node [anchor=north west][inner sep=0.75pt]    {$m$};
% Text Node
\draw (293.5,174.55) node [anchor=north west][inner sep=0.75pt]    {$x$};

\end{tikzpicture}
    \caption{There is some $x\in \CD$ such that it forms a junction with $m$ on the boundary.}
    \label{fig:x ending to give m}
\end{figure}
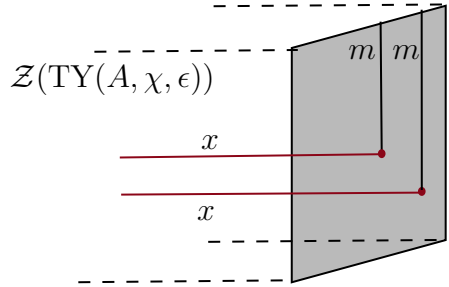

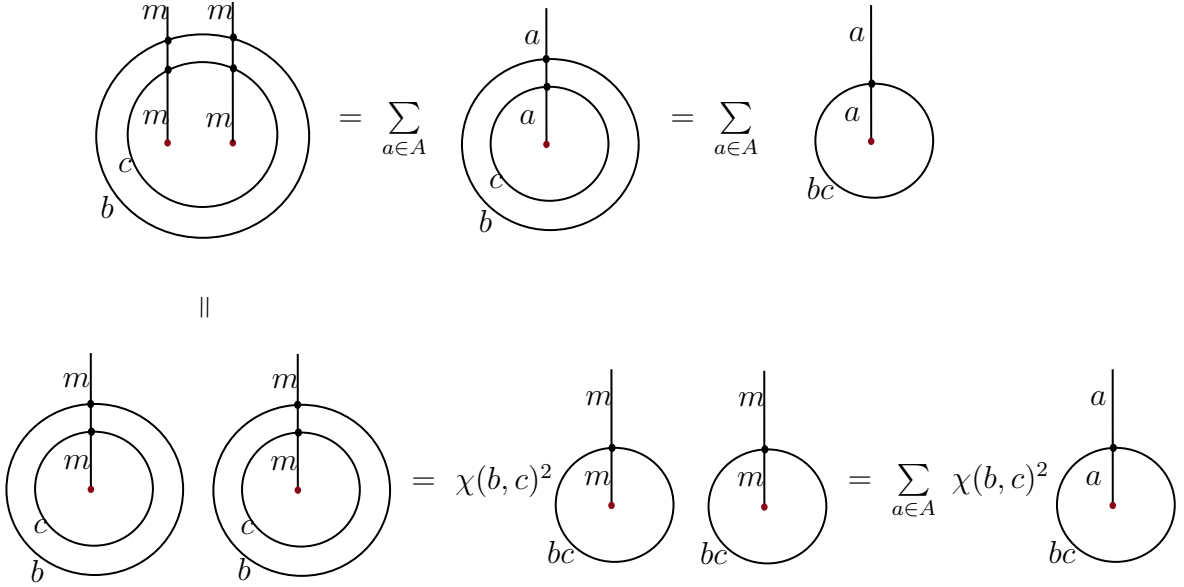
\begin{figure}[h!]
    \centering

\tikzset{every picture/.style={line width=0.75pt}} %set default line width to 0.75pt        

\begin{tikzpicture}[x=0.75pt,y=0.75pt,yscale=-0.9,xscale=0.9]
%uncomment if require: \path (0,436); %set diagram left start at 0, and has height of 436

%Straight Lines [id:da7722580120157283] 
\draw    (96.38,94.82) -- (96.33,18.45) ;
%Shape: Ellipse [id:dp21727107381925603] 
\draw  [color={rgb, 255:red, 139; green, 6; blue, 24 }  ,draw opacity=1 ][fill={rgb, 255:red, 139; green, 6; blue, 24 }  ,fill opacity=1 ] (95.13,94.59) .. controls (95.13,93.76) and (95.69,93.1) .. (96.38,93.1) .. controls (97.06,93.1) and (97.62,93.76) .. (97.62,94.59) .. controls (97.62,95.41) and (97.06,96.08) .. (96.38,96.08) .. controls (95.69,96.08) and (95.13,95.41) .. (95.13,94.59) -- cycle ;
%Shape: Ellipse [id:dp8996263007726338] 
\draw  [color={rgb, 255:red, 139; green, 6; blue, 24 }  ,draw opacity=1 ][fill={rgb, 255:red, 139; green, 6; blue, 24 }  ,fill opacity=1 ] (131.33,94.59) .. controls (131.33,93.76) and (131.89,93.1) .. (132.57,93.1) .. controls (133.26,93.1) and (133.82,93.76) .. (133.82,94.59) .. controls (133.82,95.41) and (133.26,96.08) .. (132.57,96.08) .. controls (131.89,96.08) and (131.33,95.41) .. (131.33,94.59) -- cycle ;
%Straight Lines [id:da4967662812212267] 
\draw    (132.57,93.96) -- (132.52,15.86) ;
%Shape: Ellipse [id:dp11488516713565922] 
\draw   (74.25,89.94) .. controls (74.25,67.58) and (92.83,49.45) .. (115.75,49.45) .. controls (138.67,49.45) and (157.25,67.58) .. (157.25,89.94) .. controls (157.25,112.29) and (138.67,130.42) .. (115.75,130.42) .. controls (92.83,130.42) and (74.25,112.29) .. (74.25,89.94) -- cycle ;
%Shape: Ellipse [id:dp9096719887385669] 
\draw  [color={rgb, 255:red, 0; green, 0; blue, 0 }  ,draw opacity=1 ][fill={rgb, 255:red, 0; green, 0; blue, 0 }  ,fill opacity=1 ] (131.49,52.9) .. controls (131.49,51.93) and (132.12,51.15) .. (132.89,51.15) .. controls (133.66,51.15) and (134.29,51.93) .. (134.29,52.9) .. controls (134.29,53.86) and (133.66,54.65) .. (132.89,54.65) .. controls (132.12,54.65) and (131.49,53.86) .. (131.49,52.9) -- cycle ;
%Shape: Ellipse [id:dp7403449043025166] 
\draw  [color={rgb, 255:red, 0; green, 0; blue, 0 }  ,draw opacity=1 ][fill={rgb, 255:red, 0; green, 0; blue, 0 }  ,fill opacity=1 ] (95.3,53.76) .. controls (95.3,52.8) and (95.92,52.01) .. (96.69,52.01) .. controls (97.47,52.01) and (98.09,52.8) .. (98.09,53.76) .. controls (98.09,54.72) and (97.47,55.51) .. (96.69,55.51) .. controls (95.92,55.51) and (95.3,54.72) .. (95.3,53.76) -- cycle ;
%Straight Lines [id:da6475860223307558] 
\draw    (306.51,95.68) -- (306.46,19.31) ;
%Shape: Ellipse [id:dp30485248722058955] 
\draw  [color={rgb, 255:red, 139; green, 6; blue, 24 }  ,draw opacity=1 ][fill={rgb, 255:red, 139; green, 6; blue, 24 }  ,fill opacity=1 ] (305.26,95.45) .. controls (305.26,94.62) and (305.82,93.96) .. (306.51,93.96) .. controls (307.19,93.96) and (307.75,94.62) .. (307.75,95.45) .. controls (307.75,96.27) and (307.19,96.94) .. (306.51,96.94) .. controls (305.82,96.94) and (305.26,96.27) .. (305.26,95.45) -- cycle ;
%Shape: Ellipse [id:dp5846416665955817] 
\draw   (275.55,95.1) .. controls (275.55,77.5) and (290.18,63.23) .. (308.22,63.23) .. controls (326.26,63.23) and (340.89,77.5) .. (340.89,95.1) .. controls (340.89,112.7) and (326.26,126.97) .. (308.22,126.97) .. controls (290.18,126.97) and (275.55,112.7) .. (275.55,95.1) -- cycle ;
%Shape: Ellipse [id:dp3248717236038532] 
\draw  [color={rgb, 255:red, 0; green, 0; blue, 0 }  ,draw opacity=1 ][fill={rgb, 255:red, 0; green, 0; blue, 0 }  ,fill opacity=1 ] (305.43,63.23) .. controls (305.43,62.27) and (306.05,61.49) .. (306.82,61.49) .. controls (307.6,61.49) and (308.22,62.27) .. (308.22,63.23) .. controls (308.22,64.2) and (307.6,64.98) .. (306.82,64.98) .. controls (306.05,64.98) and (305.43,64.2) .. (305.43,63.23) -- cycle ;
%Shape: Ellipse [id:dp7625974371690247] 
\draw   (56.8,90.8) .. controls (56.8,59.4) and (83.24,33.95) .. (115.85,33.95) .. controls (148.46,33.95) and (174.9,59.4) .. (174.9,90.8) .. controls (174.9,122.19) and (148.46,147.65) .. (115.85,147.65) .. controls (83.24,147.65) and (56.8,122.19) .. (56.8,90.8) -- cycle ;
%Shape: Ellipse [id:dp23089802991619612] 
\draw  [color={rgb, 255:red, 0; green, 0; blue, 0 }  ,draw opacity=1 ][fill={rgb, 255:red, 0; green, 0; blue, 0 }  ,fill opacity=1 ] (95.3,37.39) .. controls (95.3,36.43) and (95.92,35.65) .. (96.69,35.65) .. controls (97.47,35.65) and (98.09,36.43) .. (98.09,37.39) .. controls (98.09,38.36) and (97.47,39.14) .. (96.69,39.14) .. controls (95.92,39.14) and (95.3,38.36) .. (95.3,37.39) -- cycle ;
%Shape: Ellipse [id:dp4584698910158507] 
\draw  [color={rgb, 255:red, 0; green, 0; blue, 0 }  ,draw opacity=1 ][fill={rgb, 255:red, 0; green, 0; blue, 0 }  ,fill opacity=1 ] (131.49,35.67) .. controls (131.49,34.71) and (132.12,33.93) .. (132.89,33.93) .. controls (133.66,33.93) and (134.29,34.71) .. (134.29,35.67) .. controls (134.29,36.64) and (133.66,37.42) .. (132.89,37.42) .. controls (132.12,37.42) and (131.49,36.64) .. (131.49,35.67) -- cycle ;
%Shape: Ellipse [id:dp6891234245482577] 
\draw   (259.66,95.53) .. controls (259.66,69.13) and (281.6,47.73) .. (308.66,47.73) .. controls (335.73,47.73) and (357.66,69.13) .. (357.66,95.53) .. controls (357.66,121.94) and (335.73,143.34) .. (308.66,143.34) .. controls (281.6,143.34) and (259.66,121.94) .. (259.66,95.53) -- cycle ;
%Straight Lines [id:da438323697722231] 
\draw    (486.62,93.96) -- (486.57,17.58) ;
%Shape: Ellipse [id:dp5723872143232849] 
\draw  [color={rgb, 255:red, 139; green, 6; blue, 24 }  ,draw opacity=1 ][fill={rgb, 255:red, 139; green, 6; blue, 24 }  ,fill opacity=1 ] (485.38,93.73) .. controls (485.38,92.9) and (485.93,92.23) .. (486.62,92.23) .. controls (487.3,92.23) and (487.86,92.9) .. (487.86,93.73) .. controls (487.86,94.55) and (487.3,95.22) .. (486.62,95.22) .. controls (485.93,95.22) and (485.38,94.55) .. (485.38,93.73) -- cycle ;
%Shape: Ellipse [id:dp8120482992219883] 
\draw   (455.67,93.38) .. controls (455.67,75.78) and (470.29,61.51) .. (488.33,61.51) .. controls (506.37,61.51) and (521,75.78) .. (521,93.38) .. controls (521,110.98) and (506.37,125.25) .. (488.33,125.25) .. controls (470.29,125.25) and (455.67,110.98) .. (455.67,93.38) -- cycle ;
%Shape: Ellipse [id:dp14368046704374615] 
\draw  [color={rgb, 255:red, 0; green, 0; blue, 0 }  ,draw opacity=1 ][fill={rgb, 255:red, 0; green, 0; blue, 0 }  ,fill opacity=1 ] (485.54,61.51) .. controls (485.54,60.55) and (486.16,59.77) .. (486.93,59.77) .. controls (487.71,59.77) and (488.33,60.55) .. (488.33,61.51) .. controls (488.33,62.48) and (487.71,63.26) .. (486.93,63.26) .. controls (486.16,63.26) and (485.54,62.48) .. (485.54,61.51) -- cycle ;
%Shape: Ellipse [id:dp40978313216544304] 
\draw  [color={rgb, 255:red, 0; green, 0; blue, 0 }  ,draw opacity=1 ][fill={rgb, 255:red, 0; green, 0; blue, 0 }  ,fill opacity=1 ] (304.98,47.73) .. controls (304.98,46.77) and (305.61,45.98) .. (306.38,45.98) .. controls (307.15,45.98) and (307.78,46.77) .. (307.78,47.73) .. controls (307.78,48.7) and (307.15,49.48) .. (306.38,49.48) .. controls (305.61,49.48) and (304.98,48.7) .. (304.98,47.73) -- cycle ;
%Straight Lines [id:da37163999005484194] 
\draw    (53.84,288.45) -- (53.79,212.08) ;
%Shape: Ellipse [id:dp13730062060076342] 
\draw  [color={rgb, 255:red, 139; green, 6; blue, 24 }  ,draw opacity=1 ][fill={rgb, 255:red, 139; green, 6; blue, 24 }  ,fill opacity=1 ] (52.6,288.22) .. controls (52.6,287.4) and (53.16,286.73) .. (53.84,286.73) .. controls (54.53,286.73) and (55.09,287.4) .. (55.09,288.22) .. controls (55.09,289.04) and (54.53,289.71) .. (53.84,289.71) .. controls (53.16,289.71) and (52.6,289.04) .. (52.6,288.22) -- cycle ;
%Shape: Ellipse [id:dp5464420974923764] 
\draw   (22.89,287.87) .. controls (22.89,270.27) and (37.52,256.01) .. (55.56,256.01) .. controls (73.6,256.01) and (88.23,270.27) .. (88.23,287.87) .. controls (88.23,305.48) and (73.6,319.74) .. (55.56,319.74) .. controls (37.52,319.74) and (22.89,305.48) .. (22.89,287.87) -- cycle ;
%Shape: Ellipse [id:dp6097745834365118] 
\draw  [color={rgb, 255:red, 0; green, 0; blue, 0 }  ,draw opacity=1 ][fill={rgb, 255:red, 0; green, 0; blue, 0 }  ,fill opacity=1 ] (52.76,256.01) .. controls (52.76,255.04) and (53.39,254.26) .. (54.16,254.26) .. controls (54.93,254.26) and (55.56,255.04) .. (55.56,256.01) .. controls (55.56,256.97) and (54.93,257.75) .. (54.16,257.75) .. controls (53.39,257.75) and (52.76,256.97) .. (52.76,256.01) -- cycle ;
%Shape: Ellipse [id:dp8621920182779043] 
\draw   (7,288.31) .. controls (7,261.9) and (28.94,240.5) .. (56,240.5) .. controls (83.06,240.5) and (105,261.9) .. (105,288.31) .. controls (105,314.71) and (83.06,336.11) .. (56,336.11) .. controls (28.94,336.11) and (7,314.71) .. (7,288.31) -- cycle ;
%Shape: Ellipse [id:dp20695795968669795] 
\draw  [color={rgb, 255:red, 0; green, 0; blue, 0 }  ,draw opacity=1 ][fill={rgb, 255:red, 0; green, 0; blue, 0 }  ,fill opacity=1 ] (52.32,240.5) .. controls (52.32,239.54) and (52.95,238.75) .. (53.72,238.75) .. controls (54.49,238.75) and (55.12,239.54) .. (55.12,240.5) .. controls (55.12,241.47) and (54.49,242.25) .. (53.72,242.25) .. controls (52.95,242.25) and (52.32,241.47) .. (52.32,240.5) -- cycle ;
%Straight Lines [id:da27817055386494993] 
\draw    (168.62,288.92) -- (168.57,212.55) ;
%Shape: Ellipse [id:dp7946131667895016] 
\draw  [color={rgb, 255:red, 139; green, 6; blue, 24 }  ,draw opacity=1 ][fill={rgb, 255:red, 139; green, 6; blue, 24 }  ,fill opacity=1 ] (167.38,288.69) .. controls (167.38,287.87) and (167.94,287.2) .. (168.62,287.2) .. controls (169.31,287.2) and (169.86,287.87) .. (169.86,288.69) .. controls (169.86,289.52) and (169.31,290.18) .. (168.62,290.18) .. controls (167.94,290.18) and (167.38,289.52) .. (167.38,288.69) -- cycle ;
%Shape: Ellipse [id:dp7213333114882273] 
\draw   (137.67,288.35) .. controls (137.67,270.75) and (152.29,256.48) .. (170.34,256.48) .. controls (188.38,256.48) and (203,270.75) .. (203,288.35) .. controls (203,305.95) and (188.38,320.22) .. (170.34,320.22) .. controls (152.29,320.22) and (137.67,305.95) .. (137.67,288.35) -- cycle ;
%Shape: Ellipse [id:dp06441219391903141] 
\draw  [color={rgb, 255:red, 0; green, 0; blue, 0 }  ,draw opacity=1 ][fill={rgb, 255:red, 0; green, 0; blue, 0 }  ,fill opacity=1 ] (167.54,256.48) .. controls (167.54,255.51) and (168.17,254.73) .. (168.94,254.73) .. controls (169.71,254.73) and (170.34,255.51) .. (170.34,256.48) .. controls (170.34,257.44) and (169.71,258.23) .. (168.94,258.23) .. controls (168.17,258.23) and (167.54,257.44) .. (167.54,256.48) -- cycle ;
%Shape: Ellipse [id:dp9540840392762839] 
\draw   (121.78,288.78) .. controls (121.78,262.38) and (143.72,240.98) .. (170.78,240.98) .. controls (197.84,240.98) and (219.78,262.38) .. (219.78,288.78) .. controls (219.78,315.18) and (197.84,336.58) .. (170.78,336.58) .. controls (143.72,336.58) and (121.78,315.18) .. (121.78,288.78) -- cycle ;
%Shape: Ellipse [id:dp7021636423857854] 
\draw  [color={rgb, 255:red, 0; green, 0; blue, 0 }  ,draw opacity=1 ][fill={rgb, 255:red, 0; green, 0; blue, 0 }  ,fill opacity=1 ] (167.1,240.98) .. controls (167.1,240.01) and (167.73,239.23) .. (168.5,239.23) .. controls (169.27,239.23) and (169.89,240.01) .. (169.89,240.98) .. controls (169.89,241.94) and (169.27,242.72) .. (168.5,242.72) .. controls (167.73,242.72) and (167.1,241.94) .. (167.1,240.98) -- cycle ;
%Straight Lines [id:da5496790838038813] 
\draw    (342.61,297.48) -- (342.56,221.11) ;
%Shape: Ellipse [id:dp2947314204126825] 
\draw  [color={rgb, 255:red, 139; green, 6; blue, 24 }  ,draw opacity=1 ][fill={rgb, 255:red, 139; green, 6; blue, 24 }  ,fill opacity=1 ] (341.36,297.25) .. controls (341.36,296.42) and (341.92,295.76) .. (342.61,295.76) .. controls (343.29,295.76) and (343.85,296.42) .. (343.85,297.25) .. controls (343.85,298.07) and (343.29,298.74) .. (342.61,298.74) .. controls (341.92,298.74) and (341.36,298.07) .. (341.36,297.25) -- cycle ;
%Shape: Ellipse [id:dp821733083373787] 
\draw   (311.65,296.9) .. controls (311.65,279.3) and (326.28,265.03) .. (344.32,265.03) .. controls (362.36,265.03) and (376.99,279.3) .. (376.99,296.9) .. controls (376.99,314.5) and (362.36,328.77) .. (344.32,328.77) .. controls (326.28,328.77) and (311.65,314.5) .. (311.65,296.9) -- cycle ;
%Shape: Ellipse [id:dp4510695522400312] 
\draw  [color={rgb, 255:red, 0; green, 0; blue, 0 }  ,draw opacity=1 ][fill={rgb, 255:red, 0; green, 0; blue, 0 }  ,fill opacity=1 ] (341.53,265.03) .. controls (341.53,264.07) and (342.15,263.29) .. (342.92,263.29) .. controls (343.7,263.29) and (344.32,264.07) .. (344.32,265.03) .. controls (344.32,266) and (343.7,266.78) .. (342.92,266.78) .. controls (342.15,266.78) and (341.53,266) .. (341.53,265.03) -- cycle ;
%Straight Lines [id:da6536568991438094] 
\draw    (427.36,298.34) -- (427.31,221.97) ;
%Shape: Ellipse [id:dp15875899551009587] 
\draw  [color={rgb, 255:red, 139; green, 6; blue, 24 }  ,draw opacity=1 ][fill={rgb, 255:red, 139; green, 6; blue, 24 }  ,fill opacity=1 ] (426.12,298.11) .. controls (426.12,297.29) and (426.68,296.62) .. (427.36,296.62) .. controls (428.05,296.62) and (428.61,297.29) .. (428.61,298.11) .. controls (428.61,298.93) and (428.05,299.6) .. (427.36,299.6) .. controls (426.68,299.6) and (426.12,298.93) .. (426.12,298.11) -- cycle ;
%Shape: Ellipse [id:dp9368769507440707] 
\draw   (396.41,297.77) .. controls (396.41,280.16) and (411.04,265.9) .. (429.08,265.9) .. controls (447.12,265.9) and (461.75,280.16) .. (461.75,297.77) .. controls (461.75,315.37) and (447.12,329.63) .. (429.08,329.63) .. controls (411.04,329.63) and (396.41,315.37) .. (396.41,297.77) -- cycle ;
%Shape: Ellipse [id:dp8907103831165212] 
\draw  [color={rgb, 255:red, 0; green, 0; blue, 0 }  ,draw opacity=1 ][fill={rgb, 255:red, 0; green, 0; blue, 0 }  ,fill opacity=1 ] (426.28,265.9) .. controls (426.28,264.93) and (426.91,264.15) .. (427.68,264.15) .. controls (428.45,264.15) and (429.08,264.93) .. (429.08,265.9) .. controls (429.08,266.86) and (428.45,267.64) .. (427.68,267.64) .. controls (426.91,267.64) and (426.28,266.86) .. (426.28,265.9) -- cycle ;
%Straight Lines [id:da5233902085711936] 
\draw    (620.61,297.48) -- (620.56,221.11) ;
%Shape: Ellipse [id:dp5738651551007906] 
\draw  [color={rgb, 255:red, 139; green, 6; blue, 24 }  ,draw opacity=1 ][fill={rgb, 255:red, 139; green, 6; blue, 24 }  ,fill opacity=1 ] (619.36,297.25) .. controls (619.36,296.42) and (619.92,295.76) .. (620.61,295.76) .. controls (621.29,295.76) and (621.85,296.42) .. (621.85,297.25) .. controls (621.85,298.07) and (621.29,298.74) .. (620.61,298.74) .. controls (619.92,298.74) and (619.36,298.07) .. (619.36,297.25) -- cycle ;
%Shape: Ellipse [id:dp5552901698748132] 
\draw   (589.65,296.9) .. controls (589.65,279.3) and (604.28,265.03) .. (622.32,265.03) .. controls (640.36,265.03) and (654.99,279.3) .. (654.99,296.9) .. controls (654.99,314.5) and (640.36,328.77) .. (622.32,328.77) .. controls (604.28,328.77) and (589.65,314.5) .. (589.65,296.9) -- cycle ;
%Shape: Ellipse [id:dp32001921783100473] 
\draw  [color={rgb, 255:red, 0; green, 0; blue, 0 }  ,draw opacity=1 ][fill={rgb, 255:red, 0; green, 0; blue, 0 }  ,fill opacity=1 ] (619.53,265.03) .. controls (619.53,264.07) and (620.15,263.29) .. (620.92,263.29) .. controls (621.7,263.29) and (622.32,264.07) .. (622.32,265.03) .. controls (622.32,266) and (621.7,266.78) .. (620.92,266.78) .. controls (620.15,266.78) and (619.53,266) .. (619.53,265.03) -- cycle ;

% Text Node
\draw (80.44,16.01) node [anchor=north west][inner sep=0.75pt]    {$m$};
% Text Node
\draw (116.64,16.01) node [anchor=north west][inner sep=0.75pt]    {$m$};
% Text Node
\draw (292.51,29.79) node [anchor=north west][inner sep=0.75pt]    {$a$};
% Text Node
\draw (67.37,101.29) node [anchor=north west][inner sep=0.75pt]    {$c$};
% Text Node
\draw (80.81,74.58) node [anchor=north west][inner sep=0.75pt]    {$m$};
% Text Node
\draw (115.75,76.31) node [anchor=north west][inner sep=0.75pt]    {$m$};
% Text Node
\draw (289.86,74.58) node [anchor=north west][inner sep=0.75pt]    {$a$};
% Text Node
\draw (189.46,70.31) node [anchor=north west][inner sep=0.75pt]    {$=\ \sum\limits_{a\in A}$};
% Text Node
\draw (273.09,110.76) node [anchor=north west][inner sep=0.75pt]    {$c$};
% Text Node
\draw (57.51,122.34) node [anchor=north west][inner sep=0.75pt]    {$b$};
% Text Node
\draw (267.29,130.54) node [anchor=north west][inner sep=0.75pt]    {$b$};
% Text Node
\draw (472.62,28.07) node [anchor=north west][inner sep=0.75pt]    {$a$};
% Text Node
\draw (469.97,72.86) node [anchor=north west][inner sep=0.75pt]    {$a$};
% Text Node
\draw (373.57,70.31) node [anchor=north west][inner sep=0.75pt]    {$=\ \sum\limits_{a\in A}$};
% Text Node
\draw (449.34,111.73) node [anchor=north west][inner sep=0.75pt]    {$bc$};
% Text Node
\draw (36.67,221.57) node [anchor=north west][inner sep=0.75pt]    {$m$};
% Text Node
\draw (37.02,267.35) node [anchor=north west][inner sep=0.75pt]    {$m$};
% Text Node
\draw (20.42,303.53) node [anchor=north west][inner sep=0.75pt]    {$c$};
% Text Node
\draw (18.62,326.04) node [anchor=north west][inner sep=0.75pt]    {$b$};
% Text Node
\draw (152.45,222.04) node [anchor=north west][inner sep=0.75pt]    {$m$};
% Text Node
\draw (151.8,267.83) node [anchor=north west][inner sep=0.75pt]    {$m$};
% Text Node
\draw (135.2,304) node [anchor=north west][inner sep=0.75pt]    {$c$};
% Text Node
\draw (133.4,324.79) node [anchor=north west][inner sep=0.75pt]    {$b$};
% Text Node
\draw (326.43,231.59) node [anchor=north west][inner sep=0.75pt]    {$m$};
% Text Node
\draw (325.79,276.38) node [anchor=north west][inner sep=0.75pt]    {$m$};
% Text Node
\draw (305.33,315.25) node [anchor=north west][inner sep=0.75pt]    {$bc$};
% Text Node
\draw (411.19,232.46) node [anchor=north west][inner sep=0.75pt]    {$m$};
% Text Node
\draw (410.54,277.25) node [anchor=north west][inner sep=0.75pt]    {$m$};
% Text Node
\draw (390.09,316.12) node [anchor=north west][inner sep=0.75pt]    {$bc$};
% Text Node
\draw (230.04,272.45) node [anchor=north west][inner sep=0.75pt]    {$=\ \chi ( b,c)^{2}$};
% Text Node
\draw (606.43,231.59) node [anchor=north west][inner sep=0.75pt]    {$a$};
% Text Node
\draw (603.79,276.38) node [anchor=north west][inner sep=0.75pt]    {$a$};
% Text Node
\draw (583.33,315.25) node [anchor=north west][inner sep=0.75pt]    {$bc$};
% Text Node
\draw (472.04,271.45) node [anchor=north west][inner sep=0.75pt]    {$=\ \sum\limits_{a\in A} \ \chi ( b,c)^{2}$};
% Text Node
\draw (120.73,178.08) node [anchor=north west][inner sep=0.75pt]  [rotate=-90.62]  {$=$};

\end{tikzpicture}
    \caption{The lasso action of $b,c\in A$ on the two junctions of $m$ with the line $x$. The line $x$ has been suppressed in this picture for clarity. The lasso action is  simplified using the fusion rule $m\times m=\sum_{a\in A} a$ and the tube algebra multiplication in \cite{Bartsch:2025drc}.}
    \label{fig:lasso on xm junction}
\end{figure}

Suppose $\CD$ is the image of an embedding of TY$(A,\chi,\epsilon)$ in $\CZ(\text{TY}(A,\chi,\epsilon))$. Then, there exists some $x \in \CD$ such that $F(x)=m$, where $F$ is the bulk-to-boundary map to the canonical boundary $\CB_{\text{TY}(A,\chi,\epsilon)}$. Consider two copies of the line $x$ ending on $\CB_{\text{TY}(A,\chi,\epsilon)}$ as in Fig. \ref{fig:x ending to give m}.
Consider the tube algebra action of some $b,c \in A$ on the two junctions of lines $x$ with $m$ on the boundary. The multiplication of this tube algebra is computed in \cite[Section C.2]{Bartsch:2025drc}. We can simplify this by either fusing the lines first, and then acting with the lasso, or by acting with the lasso on each of the junctions, and then fusing the two lines (see Fig \ref{fig:lasso on xm junction}). 
This gives the constraint 
\be
\chi(b,c)^2=1~. 
\ee
Since, $\chi$ should be a non-degenerate symmetric bicharacter the above constraint is satisfied if and only if $A\cong \DZ_2^{M}$.

\section{Braidings on topological surfaces and lines from 3+1d SymTFT}

\label{sec:braidings on 2cat}

In spacetime dimensions higher than two, along with topological lines, we also have topological surface operators. The properties of topological surfaces and lines in 2+1d are believed to be captured by a fusion 2-category. Similar to our analysis in lower dimensions, we can ask whether a fusion 2-category can be a symmetry of a 3+1d QFT. Surface and lines can braid in 3+1d, therefore a 2+1d fusion category can be a symmetry of 3+1d QFT only if it can be promoted to a braided fusion 2-category. To determine braidings on a fusion 2-category, it will be crucial to understand the structure of 3+1d SymTFTs.

Let $\FC$ be a fusion 2-category of symmetries of a 2+1D QFT. The line operators in $\FC$ form a braided fusion category and will be denoted as $\Omega \FC$. Among the surface operators in $\FC$ some are condensation defects of these line operators. $\Sigma \Omega \FC$ will denote the fusion sub-2-category of surface operators in $\FC$ obtained from higher-gauging the line operators in $\Omega \FC$. All condensation defects admit at least one topological gapped boundary. In other words, there is at least one non-trivial interface between any condensation surface operator and the trivial surface operator. More generally, we can group surfaces into equivalence classes based on whether they admit a topological junction between them or not. These are called components of the fusion 2-category, denoted as $\FC_0$. Therefore, the number of surface operators up to condensation is $|\FC_0|$.  

The SymTFT of $\FC$ is given by the Drinfeld centre $\CZ(\FC)$ defined in \cite[Section 3]{baez1996higher}. (See also \cite{CRANS1998183}) Similar to the case of Drinfeld centre of fusion category, the objects of the Drinfeld centre  $\CZ(\FC)$ are of the form
\be
(S,e_S)
\ee
where $S \in \FC$ and $e_S$ are half-braidings
\be
e_S(X): S \times X \xrightarrow{\sim} X \times S~.
\ee
 with $X\in \FC$. The half-braiding $e_{S}$ has to satisfy several consistency conditions. Similar to the case for categories the half-braidings $e_S(X)$ are defined such that surfaces and lines in $\CZ(\FC)$ can braid consistently. Therefore, $\CZ(\FC)$ is always a braided fusion 2-category. In fact, it is expected that $\CZ(\FC)$ satisfies the remote detectability criterion which ensures that any given surface braids non-trivially with at least one line-operator. $\CZ(\FC)$ is therefore a 3+1d TQFT. The SymTFTs in 3+1d fall into two families. These follows from the classification of TQFTs in 3+1d \cite{Lan:2018vjb,Lan:2018bui,Johnson-Freyd:2020usu,Johnson-Freyd:2020twl}. A general 3+1d TQFT has point operators, line operators, surface operators and 3-dimensional membrane operators. We will assume that the 3+1d TQFT do not have any point operators except the trivial one. In such a TQFT, the membrane operators are all condensation defects of the lower dimensional operators \cite{Johnson-Freyd:2020usu}. Therefore, the TQFT is completely determined by the data of its surface and line operators. Line operators braid trivially between themselves in 3+1d. Therefore, in any 3+1d TQFT the line operators are described by a symmetric fusion category. By Deligne's theorem, such a category is of the form Rep$(G)$ or Rep$(G,z)$ where $z$ is an order two central element of $G$ \cite{deligne2002categories}. Here Rep$(G,z)$ is a symmetric fusion category which describes both bosonic and fermionic line operators. We have the following two cases:

\begin{enumerate}
	\item In a 3+1d TQFT in which all line operators are bosonic, the category of line operators is given by Rep$(G)$. The category Rep$(G)$ describes a set of bosonic line operators labelled by representations of the group $G$. Their fusion rules are given by the tensor product of representations of $G$.

Since all the bosonic line operators braid trivially, there is no anomaly and we can gauge them. This results in a 3+1d TQFT without any line operators and dual 0-form symmetry $G$ with some anomaly $\pi \in Z^4(G,U(1))$. Therefore, a 3+1d TQFT with bosonic line operators is a 3+1d Dijkgraaf-Witten (DW) theory with gauge group $G$ and DW-twist given by a 4-cocycle $\pi \in Z^4(G,U(1))$. Since all of these TQFTs admit gapped-boundaries, they are all SymTFTs. 
\item In a 3+1d TQFT with fermionic line operators, the category of line operators is given by Rep$(G,z)$. We can gauge all the bosonic line operators resulting in a 3+1d TQFT with a transparent fermion. Therefore, all 3+1d TQFT with fermionic line operators are obtained from gauging a symmetry of a 3+1d TQFT with a single non-trivial line operator which is fermionic. How many 3+1d TQFTs exist with this property? These were classified in \cite{Johnson-Freyd:2020twl}, and there are only two such 3+1d TQFTs!
\begin{itemize}
    \item The spin-$\DZ_2$ gauge theory whose dynamical gauge field is a spin structure. This TQFT has a single non-trivial fermionic line operator and a single non-trivial surface operator implementing a non-anomalous $\DZ_2$ 1-form symmetry. 
    \item A cousin of the spin-$\DZ_2$ gauge theory with the same set of line and surface operators and fusion rules, but now the $\DZ_2$ 1-form symmetry is anomalous. 
\end{itemize}
Among the two theories above, only the first one admits a gapped boundary \cite{johnson2024minimal}. Therefore, all TQFTs obtained from gaugings symmetries of the spin-$\DZ_2$ gauge theory are SymTFTs. In fact, we have 
\be
\text{spin-}\DZ_2 \text{ gauge theory }= \CZ(\Sigma \text{SVec})~.
\ee
\end{enumerate}

A fundamental result that we will use in the following is that if a fusion 2-category $\FC$ admits a braiding, then it is a subcategory of its SymTFT. The intuition behind this result is a generalization of the result for fusion categories. If $\FC$ admits a braiding, then the braiding can be used to push the surfaces and lines on the canonical gapped-boundary of $\CZ(\FC)$ corresponding to $\FC$ into the bulk. More explicitly, let $R_{S,P}$ denote the braiding of two surface operators $S$ and $P$. Then, we can choose the braiding as the half-braiding for the surface operarors giving an embedding $\CZ(\FC)$ \cite[Section 3.2]{CRANS1998183}
\be
S \mapsto (S,R_{S,\_})~,
\ee
of $\FC$ in $\CZ(\FC)$. Therefore, braidings on $\FC$ can be determined by studying the embedding of $\FC$ in $\CZ(\FC)$. 

Consider a general fusion 2-category $\FC$. As we discussed earlier, the line operators in $\FC$ form a braided fusion category $\Omega \FC$. Suppose $\FC$ acts as a symmetry of a 3+1d QFT. Since lines braid trivially in 3+1d, $\Omega \FC$ must be equivalent to Rep$(H)$ or Rep$(H,z)$. From now we will assume that there are no transparent fermionic line operators. Our discussion below can be generalized to include this case. Therefore, if $\FC$ admits braiding $\Omega \FC \cong \text{Rep}(H)$. After gauging these lines we get a new fusion 2-category $\FC'=\FC/\text{Rep}(H)$. The category $\FC'$ does not contain any non-trivial line operators. Therefore, using \cite[Theorem A]{Johnson-Freyd:2020ivj}, we find that $\FC' \cong 2\text{Vec}_G^{\pi}$ for some finite group $K$ and $\pi \in Z^4(G,U(1))$. This shows that if $\FC$ admits a braiding, then it can be obtained from gauging an $H$ 0-form symmetry in 2Vec$_G^{\pi}$. The fusion 2-categories that can be obtained from gauging symmetries of 2Vec$_G^{\pi}$ are called group-theoretical \cite{Decoppet:2023bay}. This leads to the proposition:
\begin{prop}
\label{prop: braided 2-categories are group-theoretical}
	A fusion 2-category $\FC$ admits a braiding only if it is group-theoretical.
\end{prop}
The converse of this statement is not true. For example, 2Vec$_G^{\pi}$ itself is group-theoretical. However, it does not admit a braiding when $G$ is non-abelian. This is clear because braiding requires the fusion of surfaces in 2Vec$_G^{\pi}$ to be commutative. Indeed, if 2Vec$_G^{\pi}$ admits a braiding, the group $G$ can act as a 1-form symmetry in 3+1d, which should clearly be abelian. 

From the discussion above, it is clear that if $\FC$ admits a braiding, its SymTFT is the $G$ DW theory with DW twist $\pi$, DW$(G,\pi)$. Therefore, for determining braidings on $\FC$, we have to find all possible embeddings of $\FC$ in DW$(G,\pi)$. In the following subsection, we will study the properties of the surface and line operators in DW$(G,\pi)$. 

\subsection{SymTFT for Invertible 0-form Symmetry}

\label{sec: SymTFT for G}

Let 2Vec$_G^{\pi}$ be the fusion 2-category describing a 0-form symmetry $G$ of a 2+1D QFT with anomaly $\pi \in Z^4(G,U(1))$. Its SymTFT is DW$(G,\pi)$. Since DW$(G,\pi)$ is obtained from gauging a $G$ 0-form symmetry of the trivial TQFT in 3+1d, it is clear that the genuine line operators of DW$(G,\pi)$ should implement the dual symmetry Rep$(G)$. The full set of surface operators of this TQFT can be constructed as the Drinfeld centre $\CZ(\text{2Vec}_G^{\pi})$ \cite{Kong:2019brm}. Recall that in 2+1d DW theory, the line operators are labelled as $([g],\rho_g)$. This notation arises from the fact that we have electrically charged lines $([e],\rho)$, magnetically charged lines $([g],\mathds{1})$ and dyonic lines with both electric and magnetic charges. Mathematically, this description of line operators arises from the following equivalence of 1-categories \cite{willerton2008twisted} 
\be
\mathcal{Z}(\text{Vec}_G^{\omega})\cong \oplus_{[g]} \text{Rep}(C_g,\tau_g(\omega))
\ee
where the sum above is over representatives of conjugacy classes of the group $G$, $C_g$ is the centralizer of $g$ and $\tau_g(\omega)$ is a 2-cocycle obtained from the transgression of $\omega$. The summand $\text{Rep}(C_g,\tau_g(\omega))$ is the representation category of the central extension of $C_g$ determined by $\tau_g(\omega)$. A similar result for $\mathcal{Z}(2\text{Vec}_G^{\alpha})$ was proved in \cite{Kong:2019brm}
\be
\label{2catRep}
\mathcal{Z}(2\text{Vec}_G^{\pi})\cong \oplus_{[g]} 2\text{Rep}(C_g,\tau_g(\pi))
\ee
where $2\text{Rep}(C_g,\tau_g(\pi))$ is a 2-category of right module categories over the monoidal 1-category Vec$_{C_g}^{\tau_g(\pi)}$. In other words, the simple objects of $2\text{Rep}(C_g,\tau_g(\pi))$ are the right module categories of Vec$_{C_g}^{\tau_g(\pi)}$. These are classified by the pair $(H,\psi)$ where $H$ is a subgroup of $C_g$ such that $\tau_g(\pi)|_{H}=d\psi_g$ for a 2-cochain $\psi_g$ $C^2(H,U(1))$. The surface operators of $\mathcal{Z}(2\text{Vec}_G^{\pi})$ are labelled by the simple objects of $2\text{Rep}(C_g,\tau_g(\pi))$ denoted as $S(g,H,\psi_g)$ to make their dependence on $g,H$ and $\psi_g$ explicit. The surface operator is independent of which representative we choose for the conjugacy class. Indeed, we have
\be
S(g,H,\psi_g) \cong S(h^{-1}gh,h^{-1}Hh,h^{*}(\psi_g))~,
\ee
The 2-cochain $h^{*}(\psi_g)$ is defined as
\be
\label{eq: 2cochain under conjugation}
h^{*}(\psi_g)(k,l):= \psi_g(h^{-1}kh,h^{-1}lh) ~ \tau_{hgh^{-1}}(\sigma_h{(\pi)})^{-1}~,
\ee
where $\tau_{hgh^{-1}}(\sigma_h{(\pi)})$ is a 2-cochain obtained from the transgression of the 3-cochain
\be
\sigma_h(\pi)(m,n,o):=\frac{\pi(h,m,n,o)\pi(h^{-1}mh,h^{-1}nh,h,o)}{\pi(h^{-1}mh,h,n,o)\pi(h^{-1}mh,h^{-1}nh,h^{-1}oh,h)}~.
\ee
The term $\tau_{hgh^{-1}}(\sigma_h{(\pi)})$ in \eqref{eq: 2cochain under conjugation} ensures that $\tau_{h^{-1}gh}(\pi)=d(h^{*}(\psi))$.

\subsubsection{Fusion Rules}

Given two surface operators $S(g,H,\psi_g)$ and $S(h,K,\psi_h)$, their fusion is given by \cite[Proposition 3.6]{Kong:2019brm}
\be
\label{eq:surface fusion}
S(g,H,\psi_g) \otimes S(h,K,\psi_h)= \oplus_{t \in H\backslash G / K} ~ S(g_t,H_t,\psi_{g_t})
\ee
where $g_t=t^{-1}gth$, $H_t= t^{-1}Ht \cap K$ and 
\be
\label{eq:psi after fusion}
\psi_{g_t}(k,l):=h^{*}(\psi_g)(k,l)|_{H_t} ~\psi_h(k,l)|_{H_t} ~\beta_{t^{-1}gt,h}(k,l)~.
\ee
where $h^{*}(\psi_g)$ is defined in \eqref{eq: 2cochain under conjugation} and $\beta_{t^{-1}gt,h}(k,l)$ is a 2-cochain in $C^2(H_t,U(1))$ defined as
\be
\label{eq:beta}
\beta_{t^{-1}gt,h}(k,l):= \frac{\pi(t^{-1}gt,h,k,l)\pi(t^{-1}gt,k,l,h)\pi(k,t^{-1}gt,h,l) \pi(k,l,t^{-1}gt,h)}{\pi(t^{-1}gt,k,h,l)\pi(k,t^{-1}gt,l,h)}~.
\ee

Form the fusion rules, we see that the trivial surface operator is $S(e,G,1)$, where the third entry is the trivial 2-cochain. A special class of surface operators are obtained from setting $g=e$. These surface operators are closed under fusion and form the fusion 2-category 2Rep$(G)$ \cite[Corollary 3.9]{Kong:2019brm}. These are precisely the surface operators obtained from higher-gauging various subsets of line operators in Rep$(G)$. More generally, for a fixed $g\in G$, the surface operators $S(g,H,\psi_g)$ are all related to each other by condensations.

\subsubsection{Invertible line and surface operators}

\label{sec:invertible surfaces}

The line operators in $\mathcal{Z}(2\text{Vec}_G^{\pi})$ described by Rep$(G)$ have quantum dimensions $\text{dim}(\rho)$ where $\rho$ is an irreducible representation of $G$. Therefore, line operators are non-invertible  if and only if $G$ is a non-abelian group. The invertible line operators form a group under fusion isomorphic to $G/[G,G]$ where $[G,G]$ is the commutator subgroup of $G$. 

To determine the invertible surface operators, let us consider the fusion of a surface operator with itself. Using \eqref{eq:surface fusion} we have
\be
S(g,H,\psi_g) \otimes S(g,H,\psi_g)= \oplus_{t \in H\backslash G / H} ~ S(g_t,H_t,\psi_{g_t})~.
\ee
For $S(g,H,\psi_g)$ to be invertible, this fusion should give a unique outcome. Therefore, $|H\backslash G / H|=1$. The order of a double coset is given by 
\be
|HxK|= \frac{|H||K|}{|H\cap x^{-1}Kx|}~.
\ee
Since $H\backslash G / H$ is the collection of all double cosets, for $H=K$ and $x=e$ we find that $|H\backslash G / H|=1$ if and only if $H=G$. Therefore, all invertible surface operators are of the form $S(g,G,\psi_g)$ which implies that $g$ must be in the center of $G$. Let Inv$(S)$ be the group fo invertible surface operators. This is the invertible 1-form symmetry group of the TQFT  $\mathcal{Z}(2\text{Vec}_G^{\pi})$.

\subsubsection{Non-degenerate braiding between surfaces and lines}

The number of simple line operators in $\mathcal{Z}(\text{2Vec}_G^{\pi})$ is equal to the number of irreducible representations of $G$, which is equal to the number of conjugacy classes of $G$. For every conjugacy class $[g]$, there is at least one simple surface operator $S(g,\DZ_1,1)$ (where $\DZ_1$ denotes the trivial group). Moreover, there are always at least two simple surface operators $S(e,G,1)$ and $S(e,\DZ_1,1)$ corresponding to the conjugacy class $e$. Therefore, it is clear that the number of simple surface operators is always greater than the number of simple line operators for a non-trivial group $G$. However, the surface operators $S(g,H,\psi_g)$ for a given $g$ are all related by condensations. Therefore, the number of simple surface operators up to condensation is equal to the number of simple line operators. We can put a surface $S(g,G,\psi_g)$ on a 2-sphere and consider its braiding with a line operator $\rho \in \text{Rep}(G)$ (see Fig. \ref{fig:braiding through the bulk}). 
\begin{figure}[h!]
    \centering

\tikzset{every picture/.style={line width=0.75pt}} %set default line width to 0.75pt        

\begin{tikzpicture}[x=0.75pt,y=0.75pt,yscale=-1,xscale=1]

%Shape: Ellipse [id:dp065484060477214] 
\draw  [color={rgb, 255:red, 245; green, 166; blue, 35 }  ,draw opacity=1 ][fill={rgb, 255:red, 255; green, 224; blue, 187 }  ,fill opacity=1 ] (250,145.78) .. controls (250,107.54) and (280.84,76.55) .. (318.87,76.55) .. controls (356.91,76.55) and (387.75,107.54) .. (387.75,145.78) .. controls (387.75,184.01) and (356.91,215) .. (318.87,215) .. controls (280.84,215) and (250,184.01) .. (250,145.78) -- cycle ;
%Curve Lines [id:da4846236058471258] 
\draw [color={rgb, 255:red, 245; green, 166; blue, 35 }  ,draw opacity=1 ][fill={rgb, 255:red, 248; green, 198; blue, 155 }  ,fill opacity=0.63 ]   (250,145.78) .. controls (258.16,194) and (388.21,187.65) .. (387.75,145.78) ;
%Curve Lines [id:da5991966657269066] 
\draw [color={rgb, 255:red, 245; green, 166; blue, 35 }  ,draw opacity=1 ][fill={rgb, 255:red, 248; green, 198; blue, 155 }  ,fill opacity=0.63 ] [dash pattern={on 4.5pt off 4.5pt}]  (250,145.78) .. controls (255.64,92.48) and (385.68,95.01) .. (387.75,145.78) ;
%Curve Lines [id:da7656619641873949] 
\draw [color={rgb, 255:red, 139; green, 6; blue, 24 }  ,draw opacity=1 ]   (337.24,127.88) .. controls (361.23,66.97) and (434.46,84.74) .. (426.89,157.07) ;
%Curve Lines [id:da8881762951665783] 
\draw [color={rgb, 255:red, 139; green, 6; blue, 24 }  ,draw opacity=1 ]   (341.03,188.79) .. controls (356.18,211.64) and (419,223) .. (426.89,157.07) ;
%Curve Lines [id:da33129157414497523] 
\draw [color={rgb, 255:red, 139; green, 6; blue, 24 }  ,draw opacity=1 ] [dash pattern={on 4.5pt off 4.5pt}]  (337.24,127.88) .. controls (333.46,143.11) and (322.09,164.68) .. (341.03,188.79) ;

% Text Node
\draw (175,96.4) node [anchor=north west][inner sep=0.75pt]    {$S( g,G,\psi _{g})$};
% Text Node
\draw (432,139.4) node [anchor=north west][inner sep=0.75pt]    {$\rho $};

\end{tikzpicture}
    \caption{Braiding between a surface and a line operator in 3+1d.}
    \label{fig:surface and line braiding}
\end{figure}
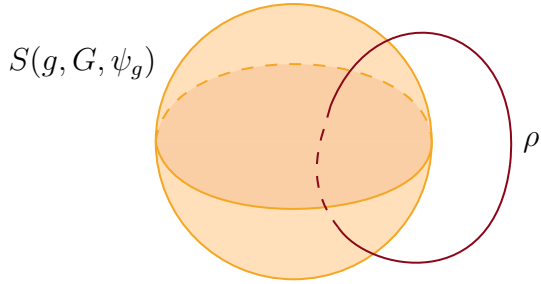

\noindent This braiding is given by 
\be
B(S(g,G,\psi_g),\rho)=\chi_{\rho}(g)~.
\ee
Note that this braiding only depends on the conjugacy class of $g$. This is because two surface operators related by condensation are equivalent when defined on a 2-sphere. Therefore, the braiding between surface operators up to condensation and line operators forms a non-degenerate matrix equal to the character table of the group $G$. This is the generalization to 3+1d of the well known fact that in a 2+1D TQFT the braiding matrix of lines is non-degenerate. 

\subsection{Braidings from embeddings}

In this section, we will describe the general procedure to determine braidings on a fusion 2-category using its 3+1d SymTFT. The SymTFT $\CZ(\FC)$ of a fusion 2-category has a canonical gapped boundary $\CB_{\FC}$ on which the surface and line operators form $\FC$. The bulk-to-boundary map is
\be
F: \CZ(\FC)\to \FC~,
\ee
determines what happens to the operators in the SymTFT when they fuse with this boundary. In particular, its action on the surface operators can be explicitly described. Recall that the simple surface operators in $\CZ(\FC)$ are given by $(S,e_S)$. The bulk-to-boundary map acts as
\be
F((S,e_S))=S~.
\ee
Note that, similarly to the case of 2+1d SymTFT, this map forgets about the half-braidings. If $\FC$ admits a braiding, then there should be an embedding
\be
\iota : \FC \to \CZ(\FC)~,
\ee
such that $F \circ \iota$ is the identity map on $\FC$. This shows that the image of the embedding $\mathfrak{D}:=\iota(\FC)$ must be such that the non-condensation surface operators and line operators in $\FD$ must not be endable on the canonical gapped boundary $\CB_{\FC}$. Let $O(\FD)$ be the set of simple topological surface operators in $\FD$ and let $O(\Omega\FD)$ be the set of simple topological line operators in the braided fusion category $\Omega\FD$. Also, let $O(L_{\FC})$ be the set of non-condensation topological surface operators in the decomposition of $L_{\FC}$ into simple surface operators. Let $E(L_{\FC})$ be the set of line operators which can end on the gapped boundary $\CB_{\FC}$. 
We require
\be 
\label{eq:embedding 2-category condition 1}
O(\FD) \cap O(L_{\FC})= \emptyset ~,~O(\Omega\FD)\cap E(L_{\FC})=\{\trl\}~,
\ee
where $\trl$ is the trivial line operator.  If the image of an embedding $\FD$ satisfies this condition, we will call it an embedding orthogonal to $L_{\FC}$. Conversely, consider a fusion sub-2-category $\FD$ of $\CZ(\FC)$ such that \eqref{eq:embedding 2-category condition 1} is satisfied. This ensures that $F|_{\FD}$ is injective. That is, no surface or line operator in $\FD$ is mapped to the trivial surface and line defect in $\FD$, respectively. Then, $F(\FD)$ is a fusion sub-2-category of $\FC$. Moreover, by choosing  $\FD$ appropriately we can ensure that $F:\FD\to \FC$ is surjective as well. 

From Proposition \ref{prop: braided 2-categories are group-theoretical}, we know that fusion 2-categories $\FC$ that admit a braiding are group-theoretical. Any group-theoretical fusion 2-category is determined by a finite group $G$, a 4-cocycle $\pi \in Z^4(G,U(1))$, a subgroup $H\subseteq G$ such that $\pi|_{H}=d \alpha$. It will be denoted $\FC(G,H,\pi,\alpha)$. It is constructed from gauging the $H$ symmetry of 2Vec$_G^{\pi}$. The condition $\pi|_{H}=d \alpha$ ensures that the subgroup $H$ is non-anomalous. The SymTFT of $\FC(G,H,\pi,\alpha)$ is $\CZ(\FC)\cong \CZ(\text{2Vec}_G^{\pi})$. Therefore, $\FC(G,H,\pi,\alpha)$ admits a braiding if and only if it can be embedded in $\CZ(\text{2Vec}_G^{\pi})$. Moreover, the embedding must be such that \eqref{eq:embedding 2-category condition 1} is satisfied, where now the Lagrangian algebra $L_{\FC(G,H,\pi,\alpha)}$ determines the gapped boundary of $\CZ(\FC)\cong \CZ(\text{2Vec}_G^{\pi})$ on which the surface and line operators form the fusion 2-category $\FC(G,H,\pi,\alpha)$. In the next section, we will use the discussion above to identity fusion 2-categories which admit a braiding. For various families of important fusion 2-categories, we will prove a necessary and sufficient condition for the existence of a braiding. 

\section{Examples}

\label{sec:examples 2cat}

\subsection{Braidings on $\Sigma \CC$}

An important class of fusion 2-categories is obtained from starting with a braided fusion category $\CC$ of line operators and constructing all condensation surface operators. We will denote this fusion 2-category $\Sigma \CC$. From Proposition \ref{prop: braided 2-categories are group-theoretical}, we know that $\Sigma \CC$ admits a braiding only if it is group-theoretical. In particular, this requires that $\CC=\text{Rep}(G)$ for some finite group $G$. The condensation surfaces of Rep$(G)$ form the fusion 2-category 2Rep$(G)$, which is known to admit a symmetric braiding. 

We can obtain the above result purely from the analysis of the SymTFT $\CZ(\Sigma\CC)$. Various properties of this 3+1d TQFT have been studied in \cite{davydov2021braided,johnson2024minimal}. $\Sigma \CC$ admits a braiding only if it can be embedded inside $\CZ(\Sigma \CC)$. In particular, this requires that we must be able to embed the line operators $\CC$ in the line operators in $\CZ(\Sigma \CC)$. From \cite[Lemma 2.16]{johnson2024minimal}, we have 
\be
\Omega \CZ(\Sigma \CC)=\CZ_2(\CC)~,
\ee
where $\CZ_2(\CC)$ is defined as
\be
\CZ_2(\CC):=\Big \langle a \in \CC | S_{ab}=\frac{d_ad_b}{D} \forall b \in \CC \Big \rangle~.
\ee
In other words, $\CZ_2(\CC)$ is the subcategory of line operators in $\CC$ which braid trivially with all other line operators. Therefore, $\CC$ can be embedded in its SymTFT if and only if $\CZ_2(\CC)=\CC$. This implies that $\Sigma \CC$ admits a braiding if and only if $\CC$ contains only transparent line operators. Therefore, $\CC \cong \text{Rep}(G)$ for some finite group $G$ and $\Sigma \CC\cong \text{2Rep}(G)$.
\begin{prop}
	The fusion 2-category $\Sigma \CC$ admits a braiding if and only if $\CC \cong \text{Rep}(G)$ for some finite group $G$.\footnote{Once again, we assume that there are no transparent fermionic line operators. The argument in this section can be straightforwardly generalized to this case.}
 
\end{prop}

\subsection{Braidings on 2Vec$^{\pi}_G$}

\label{sec:braidings on 2VecG}

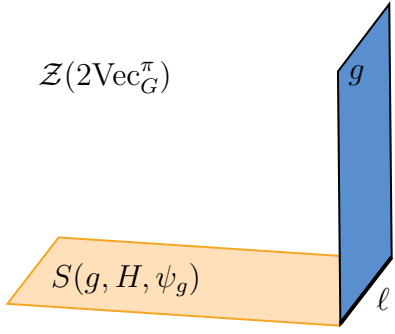
\begin{figure}
    \centering

\tikzset{every picture/.style={line width=0.75pt}} %set default line width to 0.75pt        

\begin{tikzpicture}[x=0.75pt,y=0.75pt,yscale=-1,xscale=1]

%Shape: Parallelogram [id:dp07953411453915538] 
\draw  [color={rgb, 255:red, 245; green, 166; blue, 35 }  ,draw opacity=1 ][fill={rgb, 255:red, 255; green, 224; blue, 187 }  ,fill opacity=1 ] (213.78,155.65) -- (379.01,166.59) -- (353.39,200) -- (188.16,189.05) -- cycle ;
%Shape: Parallelogram [id:dp05860148806012089] 
\draw  [fill={rgb, 255:red, 89; green, 143; blue, 206 }  ,fill opacity=1 ] (354.4,199.52) -- (353.2,72.17) -- (378.81,38.24) -- (380.01,165.59) -- cycle ;
%Straight Lines [id:da9736893829677074] 
\draw [line width=1.5]    (380.01,165.59) -- (354.39,199) ;

% Text Node
\draw (208.57,167.1) node [anchor=north west][inner sep=0.75pt]    {$S( g,H,\psi _{g})$};
% Text Node
\draw (357,67.4) node [anchor=north west][inner sep=0.75pt]    {$g$};
% Text Node
\draw (202,65.4) node [anchor=north west][inner sep=0.75pt]    {$\CZ(\text{2Vec}_G^{\pi})$};
% Text Node
\draw (371,179.4) node [anchor=north west][inner sep=0.75pt]    {$\ell $};

\end{tikzpicture}
    \caption{A bulk surface $S(g,H,\psi)$ in the embedding $\FD$ has a junction with the surface $g$ on the boundary. The category $\CC_{S(g,H,\psi_g),g}$ of line operators at this junction has a unique simple object $\ell$.}
    \label{fig:surface fusion with boundary}
\end{figure}

\begin{figure}[h!]
    \centering

\tikzset{every picture/.style={line width=0.75pt}} %set default line width to 0.75pt        

\begin{tikzpicture}[x=0.75pt,y=0.75pt,yscale=-0.9,xscale=0.9]

%Shape: Parallelogram [id:dp05860148806012089] 
\draw  [fill={rgb, 255:red, 89; green, 143; blue, 206 }  ,fill opacity=1 ] (247.25,206.24) -- (247.7,70.67) -- (430.45,39.43) -- (430,175) -- cycle ;
%Shape: Can [id:dp4304915833433014] 
\draw  [fill={rgb, 255:red, 228; green, 134; blue, 134 }  ,fill opacity=0.56 ] (244.95,155.59) -- (428.38,123.91) .. controls (440.87,121.75) and (452.69,138.96) .. (454.77,162.36) .. controls (456.86,185.75) and (448.42,206.47) .. (435.92,208.62) -- (252.5,240.31) .. controls (240,242.47) and (228.18,225.25) .. (226.1,201.86) .. controls (224.02,178.46) and (232.46,157.75) .. (244.95,155.59) .. controls (257.45,153.43) and (269.26,170.65) .. (271.35,194.04) .. controls (273.43,217.44) and (264.99,238.15) .. (252.5,240.31) ;
%Curve Lines [id:da036435886633477965] 
\draw  [dash pattern={on 4.5pt off 4.5pt}]  (433.04,209.26) .. controls (402,212) and (398.94,131.61) .. (425.41,123.62) ;
%Straight Lines [id:da08719770592425868] 
\draw [line width=1.5]    (430,175) -- (247.25,206.24) ;

% Text Node
\draw (253.77,72.98) node [anchor=north west][inner sep=0.75pt]    {$g$};
% Text Node
\draw (244,221.4) node [anchor=north west][inner sep=0.75pt]    {$h$};
% Text Node
\draw (333,192.4) node [anchor=north west][inner sep=0.75pt]    {$\ell $};
% Text Node
\draw (334,122.4) node [anchor=north west][inner sep=0.75pt]    {$u$};

\end{tikzpicture}
    \caption{The action of the tube category object $\CT_{ggh}$ on the line operator $\ell$. This induces an action of the line operator $u\in \CC_{g,h}$ on $\ell$.}
    \label{fig:tube category action}
\end{figure}
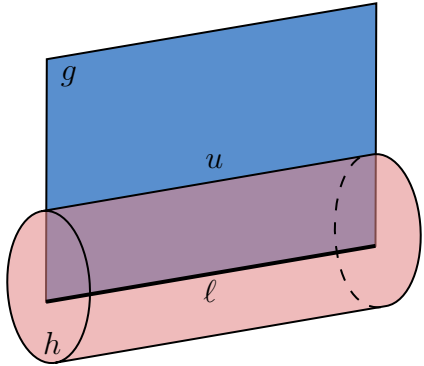

The fusion 2-category 2Vec$_G^{\pi}$ admits a braiding if and only if it can be embedded inside its SymTFT $\CZ(\text{2Vec}_G^{\pi})$. Moreover, this embedding must be orthogonal to the canonical Lagrangian algebra of the SymTFT. For $\CZ(\text{2Vec}_G^{\pi})$, this Lagrangian algebra is \cite{Zhao:2022yaw}
\be
L_{\text{2Vec}_G^{\pi}}= S(e,\DZ_1,1)~.
\ee
Recall that $S(e,\DZ_1,1)$ is the condensation surface operator obtained from higher-gauging the line operators Rep$(G)$ in the SymTFT. The Lagrangian algebra above shows that the canonical gapped boundary of $\CZ(\text{2Vec}_G^{\pi})$ is obtained from gauging its Rep$(G)$ 2-form symmetry. We studied the surface operators of this SymTFT in Section \ref{sec: SymTFT for G}. In particular, in Section \ref{sec:invertible surfaces} we learned that the invertible surface operators in $\CZ(\text{2Vec}_G^{\pi})$ are 
\be
S(g,G,\psi_g)~, 
\ee
where $g \in Z(G)$ and  $\tau_g(\pi)=d\psi_g$ for some 2-cochain $\psi_g \in C^2(G,U(1))$ form the abelian group Inv$(S)$. Consider the canonical gapped boundary $\CB_{\text{2Vec}_G^{\pi}}$ of $\CZ(\text{2Vec}_G^{\pi})$. The Lagrangian algebra corresponding to this boundary is 
\be
L_{\text{2Vec}_G^{\pi}}=S(e,G,1)~.
\ee
This encodes the fact that $\CB_{\text{2Vec}_G^{\pi}}$ is obtained from gauging the Rep$(G)$ 2-form symmetry of $\CZ(\text{2Vec}_G^{\pi})$. The bulk-to-boundary map on the invertible surfaces give
\be
F ((g,G,\psi_g))=g~.
\ee
To embed 2Vec$_G^{\pi}$ in $\CZ(\text{2Vec}_G^{\pi})$, we must at least have an embedding 
\be
\iota : G \hookrightarrow \text{Inv}(S)
\ee
as groups. Since Inv$(S)$ is abelian, it is clear that $G$ must be an abeliang group. Let $\FD:= \iota (G)$. The embedding must be such that $F(\FD)=\text{2Vec}_G^{\pi}$. Therefore, $\iota$ should be of the form
\be
g \mapsto S(g,G,\psi_g)
\ee
for some 2-cochain $\psi_g$. Therefore, we get the following theorem.
\begin{theorem}
\label{th:braidings on 2VecG}
	The fusion 2-category 2Vec$_G^{\pi}$ admits a braiding only if $G$ is abelian and $\tau_g(\pi)=d\psi_g$ for some 2-cochain $\psi_g \in C^2(G,U(1))$ for all $g\in G$.
\end{theorem}
The condition $\tau_g(\pi)=d \psi_g$  for all $g\in G$ can be understood as the vanishing of an anomaly as follows. Consider the bulk-to-boundary map $F:\CZ(\text{2Vec}_G^{\pi})\to \text{2Vec}_G^{\pi}$. A surface operator $S(g,H,\psi_g)$ in the image of the embedding $\FD$ must satisfy (see Fig. \ref{fig:surface fusion with boundary})
\be
F(S(g,H,\psi_g))=g~.
\ee
Let $\CC_{S(g,H,\pi_g),g}$ be the category of line operators at the junction of the surface operator $S(g,H,\psi_g)$ with $g$ on the gapped boundary $\CB_{\text{2Vec}_G^{\pi}}$. The above equation implies that this category must have exactly one simple object. The tube 2-algebra action on the boundary is depicted in Fig. \ref{fig:tube category action}. We will denote this tube 2-algebra element as $\mathcal{T}_{ggh}$. Consider the category $\CC_{g,h}$ of line operators at the 4-valent junction of the surfaces $g$ and $h$. The tube algebra action defines an action of the category $\CC_{g,h}$ on $\CC_{S(g,H,\pi_g),g}$. In fact, $\CC_{S(g,H,\pi_g),g}$ is a $\CC_{g,h}$-module category. The associator for the line operators in $\CC_{g,h}$ is given by the 3-cocycle $\tau_g(\pi)$ \cite[Eq. (4.80)]{Bartsch:2023wvv}. For a non-trivial 3-cocycle $\tau_g(\pi)$, a $\CC_{g,h}$-module category must has more than one simple object. Therefore, for $\CC_{S(g,H,\pi_g),g}$ to have a single simple object, $\tau_g(\pi)$ must be trivial in cohomology for all $g \in G$.

Note that for line operators, we can have braiding even for a non-trivial 3-cocycle. This is because for finite abelian groups $H^2(G,U(1))$ can be trivial even if $H^3(G,U(1))$ is non-trivial. However, $H^3(G,U(1))$ is always non-trivial. Moreover, $\tau_g(\pi)$ for all $g \in G$ cannot be trivial if $[\pi] \in H^4(G,U(1))$ is non-trivial, see Appendix \ref{app:transgression} for the argument. This implies that 2Vec$^{\pi}_G$ does not admit a braiding for a non-trivial $\pi$. If $G$ is abelian and $\pi$ is trivial in cohomology, an embedding of $2\mathrm{Vec}_{G}^{\pi}$ in $\CZ(\text{2Vec}_G^{\pi})$ is given by 
\be
g\mapsto S(g,G,1)~.
\ee
Therefore, we have the following corollary of Theorem \ref{th:braidings on 2VecG}.
\begin{corollary}
\label{cor:2VecG}
	2Vec$_G^{\pi}$ admits a braiding if and only if $G$ is abelian and $[\pi]$ is trivial in $H^4(G,U(1))$.
\end{corollary}
Note that $\pi$ is the anomaly for gauging the symmetry $G$ in 2+1d. Equivalently, if 2Vec$_G^{\pi}$ is the 1-form symmetry of a 3+1d QFT, then $\pi$ is the anomaly which obstructs higher-gauging it on a 3-manifold to construct 3-dimensional topological operators. From our discussion above, we find the following result.
\begin{corollary}
	The 1-form symmetry $G$ of a 3+1d QFT is higher-gaugeable. 
\end{corollary}
This corollary implies that a 3+1d QFT with an invertible 1-form symmetry necessarily contains 0-form symmetry implemented by the condensation defects of the 1-form symmetry. It will be interesting to determine whether these condensation defects are invertible or not.\footnote{For 2-dimensional surface operators constructed from higher-gauging line operators, the criterion for invertibility is determined in \cite{Roumpedakis:2022aik,Buican:2023bzl}.}

\subsubsection{Braidings on 2Vec$_{\DZ_2}$}

\label{sec:braidings on Z2}

The SymTFT of 2Vec$_{\DZ_2}$ is the $\DZ_2$ DW theory/3+1d toric code. The braided fusion 2-category of this theory is described explicitly in \cite{Kong:2020wmn}. We have the line operators 
\be 
\text{Rep}(\DZ_2)=\{1,\rho\}~.
\ee
There are four surface operators $S(e,\DZ_2,1),S(e,\DZ_1,1),S(g,\DZ_2,1)$ and $S(g,\DZ_1,1)$ with fusion rules
\bea
\begin{split}
S(e,\DZ_1,1) \times S(e,\DZ_1,1)&= 2 S(e,\DZ_1,1)~,~ S(g,\DZ_2,1)\times S(g,\DZ_2,1)= S(e,\DZ_2,1)~,\\
& S(g,\DZ_2,1) \times S(e,\DZ_1,1)= S(g,\DZ_1,1)~.
\end{split}
\eea
Using the fusion rules, we know that the only non-trivial invertible surface operator is $S(g,\DZ_2,1)$ which implements a $\DZ_2$ 1-form symmetry of the TQFT. Therefore, we have the embedding of 2Vec$_{\DZ_2}$ in the SymTFT 
\be
k \mapsto S(k,\DZ_2,1)~,~ k\in \DZ_2~.
\ee
Note that in the case of line operators, every embedding of a fusion category in the 2+1d SymTFT specifies a braiding on it given by the restriction of the braiding of the SymTFT on the embedding. In the case of fusion 2-categories, there are some additional subtleties. The braided fusion 2-category 2Vec$_{\DZ_2}$ describes $\DZ_2$ 1-form symmetry in a 3+1d QFT. The braiding of the order two surface operator in 2Vec$_{\DZ_2}$  captures the anomaly of the $\DZ_2$ 1-form symmetry. Since $H^5(B^2\DZ_2,U(1)) \cong \DZ_2$ the $\DZ_2$ 1-form symmetry in 3+1d admits an anomaly \cite[Theorem 35]{Wan:2018bns}\cite{Johnson-Freyd:2020twl}. However, from the above discussion, we find that at the level of objects there is a unique embedding of 2Vec$_{\DZ_2}$ in the $\DZ_2$ DW theory. This apparent discrepancy arises because both the trivial and non-trivial classes in $H^5(B^2\DZ_2,U(1))$ defines the same braided fusion 2- category \cite[Lemma 2.1.4]{Johnson-Freyd:2020twl} with an invertible surface and invertible line operator. The presence of the invertible line operator in the SymTFT leads to an identification of the two classes in $H^5(B^2\DZ_2,U(1))$ while these two classes indeed produce distinct braidings on 2Vec$_{\DZ_2}$. In fact, 2Vec$_{\DZ_2}$ equipped with these two distinct braidings arise in the two 3+1d TQFTs with a transparent order two fermionic line operator.

\subsection{Braidings on $\FC$ admitting a non-degenerate braiding}

If a fusion 2-category $\FC$ admits a braiding, we can define the braiding matrix
\be
B(S,\rho)
\ee
between the surface operator $S$ defined on a 2-sphere and a line operator linking with it. Recall that $\FC_0$ is the set of surface operators in $\FC$ up to condensation. If the braiding matrix $B(S,\rho)$ with rows labelled by the elements of $\FC_0$ and columns labelled by the simple line operators is invertible, then it is expected that it can be realized as the symmetries of a 3+1d TQFT \cite{johnson2024minimal}. In this case, the SymTFT is given by 
\be
\CZ(\FC)= \FC \times \bar \FC~.
\ee
This follows considering a TQFT that realizes $\FC$ with the trivial membrane operator. On folding the theory along this operator, we get the TQFT described by the 2-category $\FC \times \bar \FC$. The objects of this 2-category are given by pairs of surface operators $(S_1,S_2)$, $S_1,S_2 \in \FC$. The trivial membrane operator corresponds to a gapped boundary of this folded theory with a Lagrangian algebra, say $L_{\FC}$. Therefore, $\CZ(\FC) \cong \FC \times \FC$.  We can embed $\FC$ in $\FC \times \bar \FC$ in the obvious way
\be
S \hookrightarrow (S,\mathds{1})~, 
\ee
where $S \in \FC$ and $\mathds{1}$ is the trivial surface operator. Other braidings on $\FC$ can be classified by classifying other embeddings $\FD$ of $\FC$ in $\FC \times \bar \FC$ such that $\FC \cap L_{\FC}=\{\mathds{1}\}$.   
 
 \subsubsection{Other braidings on the toric code fusion 2-category}
 
 Consider the braided fusion 2-category of surface and line operators in the 3+1d DW$(\DZ_2)$/3+1d toric code described in Section \ref{sec:braidings on Z2}. It has a non-trivial order two line operator $\rho$. The surface operators up to condensation defects fall into the equivalence classes
 \be 
 \{S(e,\DZ_2,1),S(e,\DZ_1,1)\}~,~\{S(g,\DZ_2,1),S(g,\DZ_1,1)\}~.~
 \ee 
To ease the notation in the following we will denote these four surface operators as $\mathds{1},c,m,m\times c$, respectively. 
The braiding between the surfaces and lines forms the $2 \times 2$ matrix
\be
B_{\text{DW}(\DZ_2)}=\begin{pmatrix}
	1 & 1 \\
	1& -1
\end{pmatrix}~.
\ee
 The SymTFT of this category has line operators
\be
(\trl,\trl)~, (\trl,\rho)~,~(\rho,\trl)~,~(\rho,\rho)~.
\ee
The equivalence classes of surface operators up to condensation have the representatives
\be
(\mathds{1},\mathds{1})~,~(\mathds{1},m)~,~(m,\mathds{1})~,~(m,m)~.
\ee 
The braiding matrix of this theory is 
\be
B_{\text{DW}(\DZ_2)\times \text{DW}(\DZ_2)}= \begin{pmatrix}
	1  & 1 & 1 & 1\\
	1  & -1 & 1 & -1\\
	1  & 1 & -1 & -1\\
	1  & -1 & -1 & 1
\end{pmatrix}
\ee
The Lagrangian algebra object corresponding to the gapped boundary of this SymTFT with operators on it forming the DW$(\DZ_2)$ category is
\be
L_{\text{DW}(\DZ_2)}= (c,c) + (m\times c, m\times c)~.
\ee
This Lagrangian algebra corresponds to gauging the $\DZ_2$ 2-form symmetry generated by the line operator $(\rho, \rho)$ together with gauging the $\DZ_2$ 1-form symmetry generated by $(m,m)$. It is readily verified from the matrix $B_{\text{DW}(\DZ_2)\times \text{DW}(\DZ_2)}$ that this gauging gives the trivial TQFT, and therefore defines a gapped boundary. We have $O(L_{\text{DW}(\DZ_2)})=\{(m,m)\}$ and $E(L_{\text{DW}(\DZ_2)})=\{(\trl,\trl),(\rho,\rho)\}$.

Other braidings on the 3+1d DW$(\DZ_2)$ fusion 2-category are given by embedding it in the SymTFT. However, the image of an embedding $\FD$ should not contain the line operator $(\rho,\rho)$ or the surface operator $(m,m)$ to satisfy \eqref{eq:embedding 2-category condition 1}. Since the operators in the DW$(\DZ_2)$ fusion 2-category is generated by the line operator $\rho$, its condensations and the surface operator $m$, the embedding is completely specified by the embedding of $\rho$ and $m$. We have the following embeddings.
\bea
\begin{split}
&\iota_1:\rho  \mapsto (\trl,\rho)~,~ m\mapsto (\mathds{1},m) ~,\\
&\iota_2:\rho \mapsto (\trl,\rho)~,~ m\mapsto (m,\mathds{1}) ~,\\
&\iota_3:\rho \mapsto (\rho,\trl)~,~ m\mapsto (m,\mathds{1}) ~,\\
&\iota_4:\rho \mapsto (\rho,\trl)~,~ m\mapsto (\mathds{1},m) ~.\\
\end{split}
\eea
The first two embeddings are related to the last two by the 0-form symmetry of the SymTFT $\text{DW}(\DZ_2)\times \text{DW}(\DZ_2)$ which exchanges the two factors of $\text{DW}(\DZ_2)$. The embedding $\iota_1$  (or equivalently $\iota_3$) defines a braiding on the $\text{DW}(\DZ_2)$ 2-category with non-trivial braiding between $\rho$ and $m$. This is the non-degenerate braided fusion 2-category that we started with. The embedding $\iota_2$ (or equivalently $\iota_4$) defines a braiding on the $\text{DW}(\DZ_2)$ 2-category with trivial braiding between $\rho$ and $m$.

 \subsection{Braidings on Tambara-Yamagami fusion 2-category}

\label{sec:braidings on 2TY}
 
 A Tambara-Yamagami fusion 2-category TY$(A,\pi)$ is determined by a finite group $G$ which is a semi-direct (wreathe) product
\be
G=(A\times A)\rtimes B
\ee 
where $A$ is a finite abelian group, $B\cong \DZ_2$ acts on $A \times A$ by permuting the two factors, and $\pi \in Z^4(G,U(1))$ such that $[\pi]|_{A\times A}$ is trivial in $H^4(A\times A,U(1))$ \cite{Decoppet:2023bay}. The category TY$(A,\pi)$ is a group theoretical fusion 2-category which can be constructed by starting from the fusion 2-category 
\be
\text{2Vec}_G^{\pi}
\ee
and gauging the $A \times \{0\}$ subgroup of $G$. On gauging this subgroup, we get a $\DZ_2$ category of the form
\be
\mathrm{TY}(A,\pi)= \text{2Vec}_{A[1]\times A[0]} ~ \boxplus~ D~,
\ee
Let us explain the topological lines and surface operators in this category. The genuine line operators are given by 
\be
\Omega(\text{TY}(A,\pi))= \text{Rep}(A)~.
\ee
The category $ \text{2Vec}_{A[1]\times A[0]}$ contains condensation defects of the line operators                
$\text{Rep}(A)$. We will denote a condensation defect as $S_{H,\psi}$ for some subgroup $H$ of $A$ and $\psi \in Z^2(H,U(1))$. The category $ \text{2Vec}_{A[1]\times A[0]}$ also contains invertible surface operators $J_g,\;g\in A$ which form the fusion 2-category 2Vec$_A$ under fusion. All other surface operators in $ \text{2Vec}_{A[1]\times A[0]}$ are fusions of these two types of surface operators. The summand $D$ is the TY defect. The fusion rules involving $D$ are given by 
\be
\begin{split}
J_g \times D= D&~,~ S_{H,\psi} \times D= |H|~ D~,\\
D \times D &= \sum_{g\in A} X_g
\end{split}
\ee
where $X_g=J_g \times S_{\DZ_1,1}$. Since TY$(A,\pi)$ is obtained from gauging the 0-form symmetry $A\times \{0\}$ of the 2-category 2Vec$_G^{\pi}$, the SymTFT of TY$(A,\pi)$ is $\CZ(\text{2Vec}_G^{\pi})$. The bulk-to-boundary map 
\be
F: \CZ(\text{2Vec}_G^{\pi}) \to \text{2Vec}_G^{\pi}
\ee
is given by $F(S(g,H,\psi_g))=\sum_{h\in [g]} h$ up to an overall factor. Since $A\times \{0\}$ is a non-anomalous subgroup of $G$, the bulk SymTFT must have a Lagrangian algebra object
\be
L_{\text{TY}(A,\pi)}:=\sum_{S\in \CZ(\text{2Vec}_G^{\pi})} ~ S,  
\ee
corresponding to it such that the surface operators implementing the 0-forms symmetry $A\times \{0\}$ are contained in $F(L_{\text{TY}(A,\pi)})$. Therefore, $L_{\text{TY}(A,\pi)}$ contains surface operators of the form 
\be
S(g,H,\psi_g)
\ee
where the conjugacy class of $g$ contains elements of the form $(l,0,0) \in (A\times A)\rtimes B$. The object $L_{\text{TY}(A,\pi)}$ is precisely the Lagrangian algebra corresponding to the gapped boundary $\CB_{\text{TY}(A,\pi)}$ of $\CZ(\text{2Vec}_G^{\pi})$. The fusion 2-category TY$(A,\pi)$ admits a braiding if and only it can be embedded in its SymTFT orthogonally to $L_{\text{TY}(A,\pi)}$. 

\subsubsection{Emedding $J_g$ in the SymTFT}
Let $\FD$ be the image of an embedding of TY$(A,\pi)$ in $\CZ(\text{2Vec}_G^{\pi})$. In particular, $\FD$ must contain invertible surface operators whose fusion rules are isomorphic to the group $A$. These surface operators are the images of $J_g \in \text{TY}(A,\pi)$ in the SymTFT. From Section \ref{sec:invertible surfaces}, we know that $S(m,H,\psi_m)$ is invertible if and only if $m$ is in the centre of $G$ and $H=G$. We have $G=(A\times A)\rtimes \DZ_2$. We will denote the elements of this group as $(g,h,c)$ where $g,h\in A$ and $c\in\DZ_2=\{0,1\}$. The group multiplication is given by 
\be
(g_1,h_1,c_1) \times (g_2,h_2,c_2)= (g_1 + c_1h_2+(1-c_1)g_2, h_1+ c_1g_2+(1-c_1)h_2,c_1 + c_2)~,
\ee
Note that we have used additive notation for the components since they belong to abelian groups. We find
\be
Z(G)= \{(g,g,0), g\in A\}~.
\ee
Therefore, the embedding must be of the form
\be
\iota(J_g) = S((g,g,0),G,\sigma_{(g,g,0)})
\ee
This gives the consistency condition $\tau_{(g,g,0)}(\pi)=d\sigma_{(g,g,0)}$ for all $g\in A$. Moreover, the fusion of these surfaces must be isomorphic to the group $A$. Using \eqref{eq:surface fusion} we have
\be
S((g,g,0),G,\sigma_{(g,g,0)}) \times S((h,h,0),G,\sigma_{(h,h,0)})= S((g+h,g+h),G,\psi_{(g+h,g+h,0)})~,
\ee
where the 2-cocycle $\psi_{(g+h,g+h,0)}$ is 
\be
\psi_{(g+h,g+h,0)}= \sigma_{(g,g,0)} \sigma_{(h,h,0)} \beta_{(g,g,0),(h,h,0)}~.
\ee
We require 
\be
\psi_{(g+h,g+h,0)}=\sigma_{(g+h,g+h,0)} + d \chi_{(g+h,g+h)}~,
\ee
where $\chi_{(g+h,g+h)}$ is some 1-cochain in $C^1(G,U(1))$.  

\subsubsection{Embedding Rep$(A)$ and $X_a$ in the SymTFT}

The image of the embedding $\FC$ must have line operators which form the category Rep$(A)$. The line operators in the SymTFT $\CZ(\text{2Vec}_G^{\pi})$ form the category Rep$(G)$. We need an embedding of Rep$(A)$ in Rep$(G)$. All fusion subcategories of Rep$(G)$ are of the form Rep$(G/N)$ for some normal subgroup $N$ of $G$. Therefore, to embed Rep$(A)$ in Rep$(G)$ we must have $A\cong G/N_A$ for some normal subgroup $N_A$ in $G$. $G$ is a non-abelian group and $A$ is abelian. The commutator subgroup $[G,G]$ of $G$ is the smallest normal subgroup such that $G/[G,G]$ is abelian. Also, if $G/N$ is abelian, then $N$ must contain the commutator subgroup. For $G=(A\times A)\rtimes \DZ_2$ we have 
\be
[G,G]=\{(g,-g,0), g\in A\}~.
\ee
Consider the group homomorphism $\gamma : G\to A$ given by 
\be
\gamma((g,h,c))= g+h~.
\ee
The kernel of this map is $[G,G] \rtimes \DZ_2$. Therefore, $G/([G,G]\times \DZ_2)\cong A$ and the choice of the normal subgroup $N_A$ is $N_A=[G,G]\rtimes \DZ_2$. This shows that there is at least one way to embed Rep$(A)$ inside the SymTFT $\CZ(\text{2Vec}_G^{\pi})$. The embedding of the condensation defect $S_{\DZ_1,1}$ is then 
\be
S_{\DZ_1,1} \mapsto S((0,0,0),N_A,1)~.
\ee
Since $X_g=J_g \times S_{\DZ_1,1}$, the embedding of $X_g$ in the SymTFT must be
\be
X_g \mapsto S((g,g,0),G,\sigma_{(g,g,0)}) \times S((0,0,0),N_A,1)= S((g,g,0),N_A,\sigma_{(g,g,0)}|_{N_A})~,
\ee
where we have used the fact that we can always choose a basis in which the 4-cocycle $\pi$ is trivial if one or more of its arguments are the identity of the group. Then, $\beta_{(g,g,0),(0,0,0)}=1$. 

\subsubsection{Embedding $D$ in the SymTFT}

Let $S((g,h,c),H,\psi_{(g,h,c)})$ be the embedding of $D$ in the SymTFT. Using \eqref{eq:surface fusion}, we have 
\be
\label{eq: fusion of embedding of D}
S((g,h,c),H,\psi_{(g,h,c)}) \times S((g,h,c),H,\psi_{(g,h,c)}) = \sum_{t \in H\backslash G/H} S(m_t,H_t,\psi_{m_t}))~,
\ee
where $m_t=t^{-1}(g,h,c)t(g,h,c)$, $H_t=t^{-1}Ht\cap H$ and 
\be
\psi_{m_t}= t^{*}(\psi_{(g,h,c)})|_{H_t} \psi_{(g,h,c)}|_{H_t}\beta_{t^{-1}(g,h,c)t,(g,h,c)}~.
\ee
The duality defect $D$ satisfies the fusion rule
\be
D\times D=\sum_{k\in A} X_k~.
\ee
Therefore, we require
\be
\label{eq: constraint on D}
S((g,h,c),H,\psi_{(g,h,c)}) \times S((g,h,c),H,\psi_{(g,h,c)})= \sum_{k\in A} S((k,k,0),N_A,\sigma_{(k,k,0)}|_{N_A})~.
\ee
Restricting to $t=(0,0,0)$ in the sum in \eqref{eq: fusion of embedding of D}, we have 
\be
S((g+ch+(1-c)g,h+cg+(1-c)h,0),H,\psi_{(g+ch+(1-c)g,h+cg+(1-c)h,0)})~.
\ee
Imposing the constraint \eqref{eq: constraint on D}, we get $H=N_A$. Also, we require $(g+ch+(1-c)g,h+cg+(1-c)h,0)=(k,k,0)$ for some $k \in A$. Therefore, if $c=0$, then $2g=2h$, or if $c=1$, then there is no constraint on $g,h$. The element $m_t=t^{-1}(g,h,c)t(g,h,c)$ must produce distinct elements of the form $(k,k,0)$. This implies that $(g,h,c)$ must not commute with any representative of the non-trivial class in $H\backslash G/H=G/N_A$. Therefore, $N_A$ must be the centralizer of $(g,h,c)$. If $c=0$, the centralizer of $(g,h,0)$ is at least $A\times A$. However, $N_A$ has to be a normal subgroup of order $2|A|$. Hence, either $|A|=2$ or $c=1$ are the only consistent options. Since we are interested in a general abelian group $A$, we will consider the case $c=1$. Moreover, $[G,G]$ is a subgroup of $N_A$. Therefore, it must commute with $(g,h,1)$. Writing this condition down explicitly, we get
\be
(-k,k,0)(g,h,1)(k,-k,0)=(g-2k,h+2k,1)~.
\ee
This is equal to $(g,h,1)$ if and only if $2k=0$ for all $k\in A$. Therefore, $A\cong \DZ_2^{M}$ for some positive integer $M$.

In summary, for $M>1$, the embedding of $D$ in the SymTFT must be of the form 
\be 
S((g,h,1),N_A,\psi_{(g,h,1)})~,
\ee
where $N_A$ is the centralizer of $(g,h,1)$ and for $t\in G/N_A$
\be
t^{*}(\psi_{(g,h,1)})|_{N_A} \psi_{(g,h,1)}|_{N_A}\beta_{t^{-1}(g,h,1)t,(g,h,1)}= \sigma_{(k,k,0)|_{N_A}}
\ee
for some $k\in A$. 

\subsubsection{TY$(A,\pi)$ with a braiding}

Using the discussion in the preceding sections, we have the following theorem.
\begin{theorem}
\label{th: braiding on TY}
	TY$(A,\pi)$, $A\neq \DZ_2$, admits a braiding if and only if $A \cong \DZ_2^{M}$ for some  integer $M$ and $\pi$ is such that
	\begin{enumerate}
		\item there exists a choice of 2-cochains $\sigma_{(g,g,0)}$ satisfying $\tau_{(g,g,0)}(\pi)=d\sigma_{(g,g,0)}$ and 
		\be
		\sigma_{(g,g,0)} \sigma_{(h,h,0)} \beta_{(g,g,0),(h,h,0)}=\sigma(g+h,g+h,0)
		\ee
		for all $g,h \in A$.
		
		\item there exists $(g,h,1)\in G$ such that the centralizer $N_A$ of $(g,h,1)$ satisfies $G/N_A\cong A$.
		\item  there exists a 2-cochain $\psi_{(g,h,l)}$ in $C^2(N_A,U(1))$ satisfying $\tau_{(g,h,1)}(\pi)=d\psi_{(g,h,1)}$ and for $t\in G/N_A$ 
		\be
		t^{*}(\psi_{(g,h,1)})|_{N_A} \psi_{(g,h,1)}|_{N_A}\beta_{t^{-1}(g,h,1)t,(g,h,1)}= \sigma_{(k,k,0)|_{N_A}}
		\ee
		for some $k\in G$.
	\end{enumerate}
	If these conditions are satisfied, the embedding is given by 
	\bea
	\begin{split}
    \label{eq:explicit embedding}
			J_k\mapsto & S((k,k,0),G,\sigma_{(k,k,0)})~,~ X_k\mapsto S((k,k,0),N_A,\sigma_{(k,k,0)}|_{N_A})~,\\ 
			& \hspace{2.2cm} D\mapsto S((g,h,1),N_A,\psi_{(g,h,1)})~.
	\end{split}
	\eea
\end{theorem}

Note that even for $A=\DZ_2$, \eqref{eq:explicit embedding} is a valid embedding of TY$(\DZ_2,\pi)$ in the SymTFT. From now on, we will restrict our analysis to $A\cong \DZ_2^M$ as this is necessary for a braiding to exist. Let
\be
G=(A\times A) \rtimes B~,
\ee
where $A=\DZ_2^k$ and $B\cong \DZ_2$ and $B$ acts on $A \times A$ by permuting the two factors of $A$. For TY$(A,\pi)$ to admit a braiding the 4-cocycle $\pi$ must satisfy 
$\tau_{x}(\pi)=d\sigma_{x} ~ \forall ~ x \in Z(G)$. Moreover, from the definition of TY$(A,\pi)$ we know that $\pi$ must be trivial in cohomology when restricted to the subgroup $ A \times A$ of $G$. In Appendix \ref{ap:4cocycle admitting braiding}, we will show that these two conditions are only satisfied when $[\pi]$ is trivial in $H^4(G,U(1))$. Therefore, we get the following corollary of Theorem \ref{th: braiding on TY}.
\begin{corollary}
\label{cor:2TY}
    TY$(A,\pi)$ admits a braiding if and only if $A\cong \DZ_2^M$ for some integer $M$ and the 4-cocycle $[\pi]$ is trivial in $H^4(G,U(1))$.
\end{corollary}

\section{Conclusion}

In this work, we described a method to explicitly classify the anomaly of higher-form symmetries using the SymTFT. In 2+1d, this led to an explicit algorithm to compute all braidings on a fusion category $\CC$ using only its fusion ring and $F$ symbols. While the hexagon equations uses the same input, they are non-linear multivariable polynomial equations whose solutions need to be sorted into equivalence classes related by basis transformation on the fusion spaces. In our method, the braidings on $\CC$ are completely determined by the modular data of $\CZ(\CC)$ and its canonical Lagrangian algebra object, which are both basis independent.  While computing the modular data of $\CZ(\CC)$ requires the $F$-symbols in a basis, its computation can be done using a polynomial time algorithm to find the central idempotents of the tube algebra. 

In 3+1d, we used similar methods to classify braidings on higher-form symmetries. We showed that, in a sense, this problem is simpler than in 2+1d, since only those fusion 2-categories $\FC$ for which $\CZ(\FC)$ is a 3+1d DW theory admits a braiding. The 3+1d DW theories are well-studied, and we used their operator content to determine embedding of various fusion 2-categories in it.

Our work leads to various interesting questions to explore in the future:
\begin{itemize}
	\item It will be interesting to find a necessary and sufficient criterion for a general fusion 2-category to admit a braiding by studying whether they can be embedded inside the relevant 3+1d DW theory. 
	\item Our method to classify braidings applies to higher dimensions as well. It will be interesting to determine the braidings on a fusion 3-category by embedding it inside the relevant 4+1d SymTFT. In particular, it will be interesting to study braidings on Tambara-Yamagami 3-categories \cite{Bhardwaj:2024xcx}.  
	\item It will be interesting to find a basis independent quantity which determines the braiding on 3+1d DW theory (similar to modular data for MTCs). This will then enable us to compute the braidings on a fusion 2-category determined by its embeddings in the DW theory. 
    \item In \cite{Balasubramanian:2024nei}, it was shown that braided fusion categories with line operators with real spins can be completely classified. Their $F$ and $R$ symbols are related to the $F$ and $R$ symbols of certain special abelian TQFTs under gauging. It will be interesting to understand if the $F$ symbols of braided fusion categories with more general spins can be obtained by combining the methods of this paper with \cite{Balasubramanian:2024nei}. 
    
\end{itemize}

\acknowledgments

We thank Mahesh Balasubramanian, Matthew Buican, Clement Delcamp, Theo Johnson-Freyd, Adrian Padellaro, Julia Plavnik, Brandon Rayhaun, David Reutter, Eric Rowell, Yunqin Zheng for discussions and comments related to this project. R.R. gratefully acknowledges the hospitality of the University of Oxford, Simons Center for Geometry and Physics, International Centre for Mathematical Sciences, University of California Los Angeles and Kavli Institute for Theoretical Physics at UC Santa Barbara, where part of this work was carried out. P.P. thanks GGI, Mittag-Leffler Institute, BIMSA, YMSC, and SIMIS, where some work on the project was done, for hospitality.

\appendix

\section{Anomaly of line operators not closed under fusion}

\label{ap: anomaly of lines not closed under fusion}

For a line operator $A$ in a fusion category $\CC$ to be a gaugeable in 1+1d, $A$ must admit the structure of a haploid special symmetric Frobenius algebra. This is the same as requiring that $A$ admits the structure of a haploid separable algebra. This is because a haploid separable algebra admits a unique special symmetric Frobenius algebra structure.  This follows from noting that from \cite[Corollary 4.4]{kong2017semisimple} we learn that separability of $A$ implies that $A$ is semi-simple. Now, from \cite[Corollary 3.10]{Fuchs:2002cm}, a haploid algebra is symmetric for any $\epsilon \in \text{Hom}(A,\trl)$. Also, using \cite[Proposition 2 (ii)]{ostrik2003module} we know that for a semisimple $A$, 
\be
A \otimes A \xrightarrow{m} A \xrightarrow{\epsilon} \trl~.
\ee
Now, using \cite[Lemma 3.12]{Fuchs:2002cm}, we have that $A$ is a special symmetric Frobenius algebra. Moreover, using \cite[Lemma 3.7 (i)]{Fuchs:2002cm} we know that this structure is unique. Conversely, from \cite[Corollary 5.4]{kong2017semisimple}, all semi-simple algebras are separable. 

Therefore, if $A$ admits an associative algebra structure, and if it is separable, then it is non-anomalous in 1+1d.\footnote{In a unitary fusion category, the natural algebras are those which respect unitarity (see, for example, \cite{reutter2023uniqueness}). However, this is not an extra constraint as all algebras in a unitary fusion category automatically respect unitarity \cite{Carpi:2022vxj}.} In order to determine the separability of $A$, we can use the following criterion. $A=\sum_{a\in \CC} a$ in a unitary fusion category is separable if and only if \cite[Corollary 3.8]{Cong:2017ffh}.
\be
\label{eq:separableA}
n_a n_b \leq \sum_{c\in \CC} N_{ab}^c n_c  ~, \forall ~ a,b \in \CC.
\ee

So far, we discussed the conditions that $A$ must satisfy in order for it to be non-anomalous in 1+1d. If $\CC$ admits a braiding, we can think of $\CC$ as line operators in a QFT in 2+1d. Now, $A$ is a 1-forms symmetry and for it to be non-anomalous, $A$ must also be a commutative algebra. In \cite[Proposition 2.25]{Frohlich:2003hm}, we learn that a commutative Frobenius algebra is symmetric iff $\theta_A=1$. In other words, all simple objects in $A$ must be bosons. In fact, using the proof of this proposition, we learn that if $A$ is a Frobenius algebra and if $\theta_A=1$, then $A$ is a commutative algebra. Therefore, if $A$ is haploid and separable (which from our discussion above guarantees that $A$ is special symmetric Frobenius), then $A$ is commutative if and only if $\theta_{A}=1$. In summary, we get the following criterions for $A$ to be a non-anomalous line operator in 2+1d:
\begin{enumerate}
	\item $A$ admits an associative algebra structure.
	\item $A$ satisfies \eqref{eq:separableA}.
	\item $\theta_{A}=1$. 
\end{enumerate}  
Note that conditions 1 and 2 only depend on $\CC$ as a fusion category. These conditions ensure that $A$ is non-anomalous and can be gauged in 1+1d. Indeed, if $A$ satisfies these two conditions, $A$ can be higher-gauged on a 2-dimensional submanifold in 2+1d, creating a surface operator. If $A$ also satisfies condition 3, then $A$ can be gauged in 2+1d. Since $\theta_{A}$ depends on the braiding on $\CC$, it is clear that even if $A$ is higher-gaugeable, whether $A$ can be gauged in 2+1d or not depends on the braiding of $\CC$. 

\section{Transgression in the cohomology of the abelian groups}

\label{app:transgression}

In this appendix we present an argument that for any finite abelian group $G$ there are no non-trivial elements of $H^4(G;\Q/\Z)\cong H^5(G;\Z)$ such that all their transgressions are trivial. We will do this explicitly by analyzing the transgression map in terms of the basis elements. 

Any finite abelian group is of the form 
\begin{equation}
 G\cong \prod_i \Z_{N_i}.
\end{equation}
For the uniqueness of such factorization (up to permutations of the factors) one can assume that the orders of the factors are powers of prime numbers, but this will not be necessary for our analysis. Let $\alpha_i$ be the element of $H^1(G;\Z_{N_i})=\mathrm{Hom}(G,\Z_{N_i})$ corresponding to the projection $G\rightarrow \Z_{N_i}$ on the $i$-th factor, and 
\begin{equation}
    \bock: H^*(G;\Z_{N_i})\rightarrow H^{*+1}(G;\Z)
\end{equation}
be the Bockstein homomorphism. We will also use the notation $N_{ijk\ldots}:=\mathrm{GCD}(N_i,N_j,N_k,\ldots)$. In the low degrees one can choose the basis in the cohomology $H^n(G;\Q/\Z)$ as follows (see e.g. \cite{Wang:2014pma}, also cf. (\ref{HAA-generators-1})-(\ref{HAA-generators}))
 \begin{align}
      n=1:\quad   & \frac{1}{N_i}\,\alpha_i;  \\
      2:\quad  & \frac{1}{N_{ij}}\,\alpha_i\alpha_j,\;\forall i<j ;\\
            3:\quad  & \frac{1}{N_i}\,\alpha_i\bock(\alpha_i),\;\forall i,\label{type-1-3coc}\\
        & \frac{1}{N_{ij}}\,\alpha_i\bock(\alpha_j)=\frac{1}{N_{ij}}\,\alpha_j\bock(\alpha_i),\;
        \forall i<j \\
        &
        \frac{1}{N_{ijk}}\alpha_i\alpha_j\alpha_k,
        \;\forall i<j<k;\\
    4:\quad & 
    \frac{1}{N_{ij}}\,\alpha_i\alpha_j\bock(\alpha_j),\,\forall i\neq j,
    \label{type-1-4coc}
    \\
    & \frac{1}{N_{ijk}}\,\alpha_i\alpha_j\bock(\alpha_k),\;\forall 
    i<j<k, 
    \label{type-2a-4coc}
     \\
    & \frac{1}{N_{ijk}}\,\alpha_i\alpha_k\bock(\alpha_j),\;\forall 
    i<j<k, 
    \label{type-2b-4coc}
    \\
     & \frac{1}{N_{ijkl}}\,\alpha_i\alpha_j\alpha_k\alpha_l,\;
    \forall i<j<k<l.
    \label{type-3-4coc}
 \end{align}
In the formulas above, the product is the cup product induced by the natural products $\Z\times \Z_{N_i}\rightarrow \Z_{N_i}$ and $
\Z_{N_i}\times \Z_{N_j}\rightarrow \Z_{N_{ij}}$. Moreover, ``$1/N$'' is understood as the map $H^*(G;\Z_N)\rightarrow H^*(G;\Q/\Z)$ induced by the inclusion of the coefficient groups $\Z_N\hookrightarrow \Q/\Z$, $x\mapsto x/N\mod 1$.

The class $\bock(\alpha_i)$ can be represented by the following explicit integral 2-cocycle 
\begin{equation}
\begin{array}{rrcl}
     \varphi: & G^2 & \longrightarrow & \Z  \\
     & (a,b) & \longmapsto & \frac{\widetilde{a_i+b_i}-\widetilde{a_i}-\widetilde{b_i}}{N_i}
\end{array}
\end{equation}
where $g_i\in \Z_{N_i}$ is the $i$-th component of $g\in G$ and $\widetilde{\cdot}$ is a lift $\Z_{N_i}\rightarrow \Z$ (e.g. such that $\widetilde{(x\mod N_i)}=x$ for $x\in [0,N_{i}-1]$).

The action of the transgression map
\begin{equation}
    \tau_g:H^*(G;\Gamma)\longrightarrow H^{*-1}(G;\Gamma),\qquad g\in G,
\end{equation}
on the basis elements can be obtained using the fact that it satisfies the Leibniz rule (with the appropriate signs) and is natural with respect to the coefficient homomorphisms $\Gamma\mapsto \Gamma'$, and that 
\begin{equation}
    \tau_g(\alpha_i)=g_i,\qquad \tau_g(\bock(\alpha_i))=0.
\end{equation}
In particular, in degree 4 we have:
\begin{equation}
    \begin{array}{rrcl}
\tau_g: & H^4(G;\Q/\Z)& \longrightarrow & H^3(G;\Q/\Z), \\
& \frac{\alpha_i\alpha_j\alpha_k\alpha_k}{N_{ijkl}} & \longmapsto & \frac{g_i\alpha_j\alpha_k\alpha_l-
g_j\alpha_i\alpha_k\alpha_l+g_k\alpha_i\alpha_j\alpha_l-g_l\alpha_i\alpha_j\alpha_k}{N_{ijkl}} \\
& \frac{\alpha_i\alpha_j\bock(\alpha_k)}{N_{ijk}} & \longmapsto & \frac{g_i\alpha_j\bock(\alpha_k)-g_j\alpha_i\bock(\alpha_k)}{N_{ijk}} \\
& \frac{\alpha_i\alpha_j\bock(\alpha_j)}{N_{ij}} & \longmapsto & \frac{g_i\alpha_j\bock(\alpha_j)-
g_j\alpha_i\bock(\alpha_j)}{N_{ij}} \\
    \end{array}
\end{equation}

Consider an arbitrary linear combination of the basis elements in degree 4 and assume that for an arbitrary $g$ the transgression $\tau_g$ gives the trivial linear combination of the basis elements in degree 3. We want to argue that in this case all the coefficients in the original linear combinations are also zero. First, consider in the image the terms proportional to the degree 3 class $\alpha_i\alpha_j\alpha_k/N_{ijk},\;i<j<k$. Take some $p$ different from $i,j,k$ and consider $g\in G$ such that $g_r=\delta_{rp}\in \Z_{N_r}$. From the formulas above one can see that there is a single degree 4 basis element -- of type (\ref{type-3-4coc}), corresponding to the ordered sequence of $i,j,k,p$ -- that can contribute to such terms. Therefore no elements of type (\ref{type-3-4coc}) can be present in the degree 4 linear combination. Next, consider in the image the terms proportional to $\alpha_i\bock(\alpha_i)/N_{i}$. Similarly, take $p$ different from $i$ and $g\in G$ such that $g_{i}=\delta_{ip}$. Again, a single basis element -- of type (\ref{type-1-4coc}), corresponding to the unordered pair $i,p$ -- that can contribute. Therefore no elements of type (\ref{type-1-4coc}) can be present in the degree 4 linear combination. Finally, consider the terms in the image proportional to $\alpha_i\bock(\alpha_j)/N_{ij},\,i<j$. Take $p$ such that $i<p<j$ (if it exists) and $g\in G$ such that $g_r=\delta_{rp}$. The only degree 4 basis element that can contribute is of type (\ref{type-2a-4coc}), corresponding to the ordered triple $i,p,j$. Thus no elements of type (\ref{type-2a-4coc}) can be present. Next, take $p$ such that $i<j<p$. The only degree 4 basis element that can contribute is of type (\ref{type-2b-4coc}), corresponding to the ordered triple $i,j,p$. Therefore all type (\ref{type-2b-4coc}) elements must be also absent. This concludes the proof the only class in $H^4(G;\Q/\Z)$ that vanishes under all transgressions is the trivial one.

Note that a similar (much simpler) argument can be done for $H^2(G;\Q/Z)$, but not for $H^3(G;\Q/\Z)$. In particular, all elements of the form (\ref{type-1-3coc}) vanish under arbitrary transgression.

\section{Cohomology of $(\Z_2^M\times \Z_2^M)\rtimes \Z_2$}

\label{ap:4cocycle admitting braiding}

In this appendix, we present some results on the cohomology of the group 
\begin{equation}
    G=(A\times A)\rtimes \Z_2,\qquad A=\Z_2^M,
\end{equation}
which are relevant to the main part of the paper. In the semidirect product, $\Z_2$ acts by permuting two $A$ factors. 

The group $G$ contains the dihedreal group $D_8$ with multiple possible embeddings:
\begin{equation}
    D_8=\Z_4\rtimes \Z_2 \cong (\Z_2\times \Z_2)\rtimes \Z_2 
    \hookrightarrow (A\times A)\rtimes \Z_2
\end{equation}
We will use the knowledge of the group cohomology of $D_8$ and the Hochschild-Serre spectral sequence 
\begin{equation}
    E_2^{p,q}=H^p(\Z_2;H^q(A\times A;\Z))
    \Longrightarrow H^{p+q}(G;\Z)\cong H^{p+q-1}(G;\Q/\Z)
    \label{HS-2nd-page}
\end{equation}
to determine the cohomology groups of $G$.

The first step is to describe $H^*(A\times A;\Z)$ as a $\Z_2$-module.
As a group, it is known to have only elements of order two in positive degrees and therefore it is generated by the image of the Bockstein map
\begin{equation}
    \bock:H^q(A\times A;\Z_2)\twoheadrightarrow H^{q+1}(A\times A;\Z),\qquad q\geq 0,
\end{equation}
The kernel is given by the image of the first Steenrod square operation $Sq^1$. The mod 2 cohomology $H^*(A\times A;\Z_2)\cong H^*(A;\Z_2)^{\otimes 2}$ also has $\Z_2$-module structure and the map is equivariant with respect to the $\Z_2$ action permitting the two factors. The mod 2 cohomology is a non-trivial $\Z_2$-module, with $\Z_2$ action induced by the permuting the two copies of $A$. Explicitly, we have:
\begin{equation}
    H^*(A\times A;\Z_2)=\Z_2[\alpha_1,\ldots,\alpha_M,\beta_1,\ldots,\beta_M]
\end{equation}
where $\alpha_{i}$ and $\beta_{i}$ are the basis elements of $H^1(A;\Z_2)\cong \mathrm{Hom}(A,\Z_2)$ for the two different copies of $A$, corresponding to the projection $A=\Z_2^{M}\rightarrow \Z_2$ on the $i$-th component. The non-trivial element of $\Z_2$ then acts as the following automorphism of $H^*(A\times A;\Z_2)$:
\begin{equation}
    \alpha_i\longleftrightarrow \beta_i.
    \label{Z2-auto-coh-generators}
\end{equation}

The basis in the groups $H^{q}(A\times A;\Z)$ can be chosen as follows in low degrees:
 \begin{align}
      q=0:\quad   & 1; \label{HAA-generators-1} \\
      1:\quad  & - \\
      2:\quad  & \bock(\alpha_i),\, \bock(\beta_i),\;\forall i; \\
      3:\quad  & \bock(\alpha_i\beta_j),\, \forall i,j,
      \\
      & \bock(\alpha_i\alpha_j), \bock(\beta_i\beta_j),\;\forall i<j ;\\
            4:\quad  & \bock(\alpha_i^3),\bock(\beta_i^3),\;\forall i,\\
        & \bock(\alpha_i\alpha_j^2),\bock(\beta_i\beta_j^2),\;
        \forall i<j \\
        &  \bock(\alpha_i\beta_j^2)=\bock(\beta_j\alpha_i^2),\; \forall i,j,\\
        &
        \bock(\alpha_i\alpha_j\alpha_k),
        \bock(\beta_i\beta_j\beta_k),\;\forall i<j<k,
        \\
        &
    \bock(\alpha_i\beta_j\beta_k), \bock(\beta_i\alpha_j\alpha_k),\;\forall i,\forall j<k; \\
    5:\quad & 
    \bock(\alpha_i\alpha_j^3),\bock(\beta_i\beta_j^3),\forall i\neq j,
    \\
    &
    \bock(\alpha_i^3\beta_j),\bock(\beta_i^3\alpha_j),\forall i,j, \\
    & \bock(\alpha_i^2\alpha_j\alpha_k),
    \bock(\beta_i^2\beta_j\beta_k),\;\forall 
    i<k,j<k,i\neq j, \\
    &
    \bock(\alpha_i^2\alpha_j\beta_k),
    \bock(\beta_i^2\beta_j\alpha_k),\;\forall 
    i\neq j,
    \\
    & \bock(\alpha_i\alpha_j\alpha_k\alpha_l),
    \bock(\beta_i\beta_j\beta_k\beta_l),\;
    \forall i<j<k<l,
    \\
    & \bock(\alpha_i\alpha_j\alpha_k\beta_l),
    \bock(\beta_i\beta_j\beta_k\alpha_l),\;
    \forall i<j<k,\,\forall l,
    \\
    & \bock(\alpha_i\alpha_j\beta_k\beta_l),\;
    \forall i<j,\, k<l,
    \label{HAA-generators}
 \end{align}
where each $\bock(\ldots)$ generates a $\Z_2$ subgroup.

In each positive degree, the cohomology $H^*(A\times A;\Z_2)$ can be split into a direct sum of irreducible $\Z_2$-modules isomorphic either to $\Z_2$ with the trivial $\Z_2$ action, or to the group ring $\Z_2[\Z_2]\cong \Z_2\oplus \Z_2$, with the $\Z_2$ action permuting two generators. Such a splitting can be done consistently with the choice of the basis above. The trivial modules are generated by the basis elements invariant under the exchange $\alpha_i\leftrightarrow\beta_i$: $\bock(\alpha_i\beta_i)$, $\bock(\alpha_i\beta_i^2)=\bock(\beta_i\alpha_i^2)$, $\bock(\alpha_i\alpha_j\beta_i\beta_j)$.

To determine the entries of the second page of the spectral sequence (\ref{HS-2nd-page}) one can then use the knowledge of the cohomology of $\Z_2$ group with coefficients in a general $\Z_2$-module (see e.g. \cite{weibel1994introduction}). For the trivial module $\Z_2$, we have
\begin{equation}
    H^*(\Z_2;\Z_2)\cong\Z_2[\gamma]
\end{equation}
where $\gamma$ has degree one. For the non-trivial module $\Z_2[\Z_2]$, the cohomology is supported in degree zero:
\begin{equation}
    H^*(\Z_2;\Z_2[\Z_2])\cong \Z_2.
\end{equation}
with the generator corresponding to the $\Z_2$-invariant subspace of the module. And for the $\Z$ coefficients with the trivial action, as before, we have
\begin{equation}
    H^*(\Z_2;\Z)=\Z1\oplus \Z_2\gamma^2
    \oplus \Z_2\gamma^4\oplus \Z_2\gamma^6\oplus \ldots
\end{equation}
where we identified $\bock(\gamma)$ with $Sq^1(\gamma)=\gamma^2$ under the mod 2 reduction.

Therefore the entries on the second page $E_2^{*,*}$ have the following generators:

\begin{center}
\scriptsize
\begin{tikzpicture}
  \matrix (m) [matrix of math nodes,
    nodes in empty cells,nodes={minimum width=5ex,
    minimum height=5ex,outer sep=-5pt},
    column sep=1ex,row sep=1ex]{
    5 & \begin{array}{c}
\Z_2\bock(\alpha_i\alpha_j\beta_i\beta_j),\,i<j, \\
     \Z_2\bock(\alpha_i^3\beta_j+\beta_i^3\alpha_j),\\
    \ldots
    \end{array} &
     \Z_2 \gamma\bock(\alpha_i\alpha_j\beta_i\beta_j),\,i<j & \ldots
     \\
    4 & \begin{array}{c}
    
    \Z_2\bock(\alpha_i\beta_i^2), \\
    \Z_2\bock(\alpha_i\beta_j^2+\beta_i\alpha_j^2),\,i\neq j,\\
    \Z_2\bock(\alpha_i^3+\beta_i^3),
    \\
    \ldots
    \end{array} &
     \Z_2\gamma\bock(\alpha_i\beta_i^2) &
    \Z_2\gamma^2\bock(\alpha_i\beta_i^2) &
    \Z_2\gamma^3\bock(\alpha_i\beta_i^2) &
    \ldots
     \\
    3 & \begin{array}{c}
        \Z_2\bock(\alpha_i\beta_i), \\
    \Z_2\bock(\alpha_i\beta_j+\beta_i\alpha_j),\,i\neq j,
    \\
    \Z_2\bock(\alpha_i\alpha_j+\beta_i\beta_j)
    \end{array} &
     \Z_2\gamma\bock(\alpha_i\beta_i) &
    \Z_2\gamma^2\bock(\alpha_i\beta_i) &
    \Z_2\gamma^3\bock(\alpha_i\beta_i) &
    \Z_2\gamma^4\bock(\alpha_i\beta_i) &
    \ldots
     \\
          2      & \Z_2\bock(\alpha_i+\beta_i) & 
          0 & 0 & 0 & 0 & 0
          \\
          1     &  0 &  0  & 0 & 0 & 0 & 0  \\
          0     &  \Z1  & 0 &  \Z_2\gamma^2  & 0 &
          \Z_2\gamma^4 & 0 \\
    \quad\strut &   0  &  1  &  2  & 3 & 4 & 5 \strut \\};
  \draw[->] (m-2-3.south east) -- (m-3-5.north west) node[midway,above] {$d_2$};
\draw[thick] (m-1-1.north east) -- (m-7-1.east) ;
\draw[thick] (m-7-1.north) -- (m-7-7.north east) ;
\end{tikzpicture}  
\end{center}
The 0-th column, $E_2^{0,*}=H^0(\Z_2,H^*(A\times A;\Z))=\left(H^*(A\times A;\Z)\right)^{\Z_2}$, is generated by the $\Z_2$-invariant combinations of the generators listed in (\ref{HAA-generators-1})-(\ref{HAA-generators}). In the degrees sufficient to determine $H^n(G;\Z)$ for $n\leq 5$, the differentials $d_r: E^{p,q}_r\rightarrow E^{p+r,q-r+1}_r$ and the extensions in the spectral sequence can be fixed using the knowledge of the cohomology of the dihedral group $D_8$. Namely, one can consider the following commutative diagrams with the rows being short exact sequences:
\begin{equation}
    \begin{tikzcd}
0 \ar[r] &  A\times A \ar[r] & G \ar[r] &  \Z_2 \ar[r] &  0 \\
0 \ar[r] &  \Z_2\times \Z_2 \ar[r] \ar[u,hook] & D_8 \ar[r] \ar[u,hook] &  \Z_2 \ar[r] \ar[u,"\cong"] &  0. \\
\end{tikzcd}
\end{equation}
The cohomology ring of $D_8$ is known \cite{handel1993products}:
\begin{equation}
H^*(D_8,\Z)=\Z[\ra_2,\rb_2,\rc_3,\rd_4]/(2\ra_2,2\rb_2,2\rc_3,4\rd_4,\rb_2^2+\ra_2\rb_2,\rc_3^2+\ra_2\rd_4)    
\end{equation}
where the index of the generator equals to its degree. Explicitly, in low degrees we have:
\begin{equation}
    H^n(D_8;\Z)=\left\{
\begin{array}{rl}
   \Z1,  & n=0, \\
   0, & n=1, \\
   \Z_2\ra_2\oplus \Z_2\rb_2,  & n=2, \\
    \Z_2\rc_2, & n=3, \\
   \Z_4\rd_4\oplus \Z_2\ra_2^2\oplus \Z_2\rb_2^2, & n=4, \\
   \Z_2\ra_2\rc_3\oplus \Z_2\rb_2\rc_3, & n=5,\\
   \ldots &
\end{array}
    \right.
\end{equation}
Using the naturality property and the product structure on the Hochschild-Serre spectral sequence, one can then conclude that, in the relevant degrees, the only non-trivial differential is $d_2$, which takes $\gamma\bock(\alpha_i\beta_i^2)$ to $ \gamma^3\bock(\alpha_i\beta_i)$ with all other generators sent to zero. Moreover, the only non-trivial extension is the extension of $\Z_2\bock(\alpha_i\beta_i^2)$ by $\Z_2\gamma\bock(\alpha_i\beta_i)$, which gives a $\Z_4$ subgroup of $H^4(G;\Z)$.
We then conclude that
\begin{equation}
    H^4(G;\Q/\Z)\cong H^5(G;\Z)=\left(H^5(A\times A;\Z)\right)^{\Z_2}
    \oplus \bigoplus_{i=1}^M \Z_2\gamma^2\bock(\alpha_i\beta_i)
\end{equation}

In the context of Tambara-Yamagami fusion 2-categories we are interested in the subgroup that trivializes under the restriction to $A\times A\subset G$. This subgroup is generated by the order 2 elements $\gamma^2\bock(\alpha_i\beta_i)$. Its representative $\Q/\Z$-valued 4-cocycle $\pi_i:G^4\rightarrow \Q/\Z$ can be chosen of the form
\begin{equation}
    \pi_i(a,b,c,d)=a_3b_3\phi_i(c,d)\mod 1
\end{equation}
 where $g_3$ genotes the image of $g\in G$ under the natural projection $G=(A\times A)\rtimes \Z_2\rightarrow \Z_2$, and $\phi$ is a $\Q/\Z$-valued 2-cocycle corresponding to the generator $\bock(\alpha_i\beta_i)\in H^3(G;\Z)\cong H^2(G;\Q/\Z)$.

More explicitly, let $g^{(i)},\,i=1,\ldots,M$ denote the image of $g\in G$ under the projection $G=(A\times A)\rtimes \Z_2\rightarrow (\Z_2\times \Z_2)\rtimes \Z_2=D_8$ where the $i$-th generator in each of the two copies of $A=\Z_2^M$ is sent to the generator to the generator of the corresponding copy of $\Z_2$. Therefore, up to a coboundary, we can assume that
\begin{equation}
    \phi_i(c,d)=\frac{1}{2}\,\phi(c^{(i)},d^{(i)})\mod 1
\end{equation}
where $\phi$ is a $\Z_2$-valued 2-cocycle on $(\Z_2\times \Z_2)\rtimes \Z_2=D_8$ such that its restriction to the $\Z_2\times \Z_2$ subgroup is given by the cup-product of the projection maps to the $\Z_2$ factors:
\begin{equation}
    \phi|_{\Z_2\times\Z_2}(c,d)=c_1d_2\mod 2
    \label{app:2-cocycle-reduction}
\end{equation}
An explicit expression for such a 2-cocycle can be found by considering a corresponding central extension of $D_8$. The result is the following:
\begin{equation}
\phi(a,b)=(1+a_3)a_1b_2+a_3(b_1+b_2+b_1(a_1+b_2))
\end{equation}
where we use the notation $g=((g_1,g_2),g_3)\in (\Z_2\times \Z_2)\rtimes \Z_2$. Alternatively, one can directly check that the coboundary of the above function $G^2\rightarrow \Z_2$ is zero and that the condition (\ref{app:2-cocycle-reduction}) is satisfied.

In the main text we are interested in the transgression of the 4-cocyles on $G$ to the 3-cocycles, with respect to the central elements $((\alpha,\alpha),0)\in (A\times A)\rtimes \Z_2$. Explicitly, the transgression reads
\begin{multline}
    (\tau_{((\alpha,\alpha),0)}\pi_i)
    (a,b,c)=a_3b_3(\phi_i(((\alpha,\alpha),0),c)+\phi_i(c,((\alpha,\alpha),0))) =\\
    \frac{1}{2}\,\alpha_i a_3b_3(c_1^{(i)}+c_2^{(i)}+c_3)\mod 1.
\end{multline}
For $\alpha_i\neq 0$ this is a non-trivial 3-cocycle on $G$, since, in particular, it restricts to a non-trivial 3-cocycle on the $\Z_2$ subgroup of the elements of the form $((0,0),*)$.

Therefore we conclude that an arbitrary non-trivial 4-cocycle on $G=(A\times A)\rtimes \Z_2$ either does not trivialize when restricted to $A\times A\subset G$ or its transgression with respect to a non-zero central element does not trivialize.

\newpage
\bibliography{refs}

\end{document}